\documentclass[pdflatex,sn-mathphys-num]{sn-jnl}

\usepackage{graphicx}%
\usepackage{multirow}%
\usepackage{amsmath,amssymb,amsfonts}%
\usepackage{amsthm}%
\usepackage{mathrsfs}%
\usepackage[title]{appendix}%
\usepackage{xcolor}%
\usepackage{textcomp}%
\usepackage{manyfoot}%
\usepackage{booktabs}%
\usepackage{algorithm}%
\usepackage{algorithmicx}%
\usepackage{algpseudocode}%
\usepackage{listings}%
\usepackage{placeins}
\usepackage{subcaption}
\usepackage{longtable}

\theoremstyle{thmstyleone}%

\theoremstyle{thmstyletwo}%

\theoremstyle{thmstylethree}%

\begin{document}

\title[Article Title]{Spectroscopy of $cc\bar{c}\bar{c}$ and $ss\bar{c}\bar{c}$ Tetraquarks within the Framework of Regge Phenomenology}


\author*[1]{\fnm{Vandan} \sur{Patel}}\email{vandankp12998@gmail.com}

\author[1]{\fnm{Juhi} \sur{Oudichhya}}\email{juhioudichhya01234@gmail.com}
\equalcont{These authors contributed equally to this work.}

\author[1]{\fnm{Ajay Kumar} \sur{Rai}}\email{raiajayk@gmail.com}
\equalcont{These authors contributed equally to this work.}

\affil*[1]{\orgdiv{Department of Physics}, \orgname{Sardar Vallabhbhai National Institute of Technology}, \orgaddress{\city{Surat}, \postcode{395007}, \state{Gujarat}, \country{India}}}



\abstract{	In this work, we investigate the mass spectra of all-charm ($cc\bar{c}\bar{c}$) and doubly strange- doubly charm ($ss\bar{c}\bar{c}$) tetraquark states using the framework of Regge phenomenology. Employing a quasi-linear Regge trajectory ansatz, we derive linear and quadratic mass inequalities for hadrons, which provide constraints on the masses of tetraquark states. We estimate the range of ground state masses of $cc\bar{c}\bar{c}$ tetraquarks and determine the Regge slope parameters by fitting the corresponding $(J, M^2)$ trajectories. These parameters are then utilized to predict the mass spectra of orbital excited states of both $cc\bar{c}\bar{c}$ and $ss\bar{c}\bar{c}$ systems in the $(J, M^2)$ plane. Furthermore, we extend our analysis to radial excitations by exploring Regge trajectories in the $(n, M^2)$ plane. The obtained mass predictions are compared with existing theoretical results from various models. Additionally, we discuss the possible identification of the experimentally observed $\psi(4660)$ and $\chi_{c0}(4700)$ resonances as tetraquark candidates. The results presented in this study offer useful benchmarks for future experimental investigations and may assist in the spin-parity assignment of exotic hadronic states. Our findings contribute to a deeper understanding of multiquark dynamics and the spectroscopy of exotic hadrons within the framework of Quantum Chromodynamics.}


\maketitle

\section{Introduction}\label{sec1}

	The quark model, originally introduced independently by Gell-Mann and Zweig in 1964~\cite{ref1,Zweig:1964jf}, provided a revolutionary foundation for classifying and analyzing hadrons based on their elementary building blocks---quarks. Over the past few years, a wide range of hadronic bound states have been observed at various experimental facilities, including LHCb~\cite{Sigma_b(6097),Cascade_b(6227),Omega_b2020,Cascade_b(6333)}, Belle~\cite{Belle2010,ref4}, BESIII~\cite{BESIII2019,BESIII2020,Exp3,Exp4}, and J-PARC~\cite{K. Aoki2021}. These discoveries have been supported by several theoretical approaches that have predicted their mass spectra and other physical properties~\cite{ref5,ref6,ref9}.
	
	Although Quantum Chromodynamics (QCD) is the fundamental theory governing strong interactions, it does not confine the hadronic spectrum to only mesons and baryons. Rather, QCD permits more complex configurations, including tetraquarks ($qq\bar{q}\bar{q}$), pentaquarks ($qqqq\bar{q}$), hybrid mesons (consisting of a quark--antiquark pair with excited gluonic fields), and glueballs (composed entirely of gluons). These exotic states, once purely theoretical, have gained substantial experimental backing over the last twenty years.
	
	Remarkable progress has been made in the discovery of exotic hadrons, particularly tetraquarks and pentaquarks, beginning with the detection of the $X(3872)$ state by the Belle Collaboration in 2003~\cite{Choi:2003ue}. Subsequent findings have included a variety of non-conventional states, such as $T_{c\bar{c}1}(3900)$~\cite{BESIII:2020oph}, $T_{b\bar{b}1}(10610)$~\cite{Belle:2013urd}, $X(4140)$~\cite{PhysRevLett.118.022003}, and the pentaquark candidates $P_c(4380)^+$ and $P_c(4450)^+$ reported by the LHCb Collaboration~\cite{Aaij:2015tga}.
	
	In 2009, the CDF Collaboration reported the discovery of the $X(4140)$ resonance with a measured mass of $M = 4143.0 \pm 2.9 \pm 1.2$~MeV and a width of $\Gamma = 11.7^{+8.4}_{-6.7} \pm 3.7$~MeV in the $B^+ \rightarrow J/\psi \phi K^+$ decay channel~\cite{Aaltonen:2009tz}. In the following years, another structure, the $X(4100)$, was identified by several other experiments including LHCb, D$\phi$, CMS, and BABAR~\cite{Abazov:2018bun,Sirunyan:2020qir,Aaij:2016iza}. Additionally, the $X(4274)$ resonance was observed in 2011 by the CDF Collaboration with a mass of $M = 4274.4 \pm 1.9$~MeV and a width of $\Gamma = 32.3 \pm 7.6$~MeV, also in the $B^+ \rightarrow J/\psi \phi K^+$ channel, and with a significance of $3.1\sigma$~\cite{CDF:2017hse}. The LHCb Collaboration later confirmed both the $X(4140)$ and $X(4274)$ states and determined their quantum numbers to be $J^{PC} = 1^{++}$~\cite{PhysRevLett.118.022003,LHCb:2016nsl}. The discovery of these states has greatly stimulated ongoing investigations into the properties and underlying structure of exotic hadrons, especially tetraquarks, from both experimental and theoretical viewpoints. Also, in very recent times the work has been done on tetraquarks's different properties like mass spectra and decay characteristics \cite{Lodha:2024light}.
	
	Analyzing the mass spectra of this exotic particles including tetraquarks offers crucial insights into the dynamics of the strong force as governed by Quantum Chromodynamics (QCD)~\cite{ref28}. Such investigations contribute to our understanding of non-perturbative aspects of QCD, which are essential for a complete description of hadronic systems. Furthermore, identifying and studying tetraquark states aids in exploring the mechanisms of color confinement and the significance of color charge in QCD~\cite{ref29}. Theoretical approaches to tetraquark studies span various methods, including lattice QCD, QCD sum rules, effective field theories, and phenomenological frameworks such as the quark model and the diquark-antidiquark model. Among these, lattice QCD stands out as a rigorous tool based on first principles, utilizing a discrete spacetime lattice to simulate QCD and predict tetraquark properties~\cite{ref36}. Phenomenological models serve as valuable tools for gaining qualitative understanding and can be fine-tuned using experimental observations to forecast the tetraquark mass spectra~\cite{ref38}. Within the quark model framework, tetraquarks are interpreted as bound systems composed of quarks and antiquarks, similar to the treatment of mesons and baryons. To describe the interaction among quarks, effective potentials-such as the Cornell potential-are frequently employed~\cite{ref59}.
	
	Although tetraquark systems containing charm or bottom quarks alongside light quarks have received significant attention, a particularly intriguing subset involves fully heavy configurations such as all-charm ($cc\bar{c}\bar{c}$) and all-bottom ($bb\bar{b}\bar{b}$) tetraquarks. These systems have been the subject of detailed investigations using various theoretical methods, including potential models~\cite{Debastiani:2017msn,Bai:2016int}, QCD sum rules~\cite{Chen:2016jxd}, lattice QCD~\cite{Francis:2016hui}, and the diquark-antidiquark picture~\cite{Karliner:2016zzc}. Among these, the all-charm tetraquark stands out due to its distinct composition and potentially narrow decay widths. Experimental efforts by collaborations such as LHCb and CMS~\cite{LHCb:2020bwg} have explored the possible existence of $cc\bar{c}\bar{c}$ states, with signals near 6.9~GeV, though no definitive discovery has been made so far.

	Alongside the all-heavy sector, another promising domain is that of heavy--strange tetraquarks, particularly the strange-charm systems. These combine heavy charm quarks with strange quarks, offering a unique interplay between SU(3) flavor symmetry breaking and heavy-quark dynamics. Theoretical investigations of $cs\bar{c}\bar{s}$ states have been done using potential model \cite{Tiwari:2023}. The potential identification of such states with known resonances would provide critical insights into multiquark dynamics.
	
	In this study, we investigate the mass spectra of $cc\bar{c}\bar{c}$ and $ss\bar{c}\bar{c}$ tetraquark systems within the framework of Regge phenomenology. Building upon the quasi-linear Regge trajectory approach, Wei et al.~\cite{ref39,Wei:2016jyk} established key mass relations for hadrons, such as quadratic mass equalities and both linear and quadratic mass inequalities. Motivated by their work, we apply a similar methodology and extend it to derive mass inequalities for excited tetraquark states, assuming linear Regge trajectories. We specifically analyze the connections between Regge slopes, intercepts, and tetraquark masses in the $(J, M^2)$ and $(n, M^2)$ planes, which allows us to estimate the mass ranges for ground and excited states with various spin-parity configurations.
	
	We examine the $0^+$, $1^+$, and $2^+$ Regge trajectories for the $cc\bar{c}\bar{c}$ and $ss\bar{c}\bar{c}$ systems to extract the slope and intercept parameters. These parameters are subsequently used to compute the mass spectra for both $cc\bar{c}\bar{c}$ and $ss\bar{c}\bar{c}$ tetraquarks. Additionally, we explore radial excitations through $(n, M^2)$ trajectories to make predictions for the excited states in $(n,M^2)$ plane also. Our results aim to support ongoing efforts in identifying and classifying multiquark hadrons, offering important benchmarks for future experimental investigations.
	
	The paper is organized as follows. Section II presents the theoretical foundation of Regge theory. In Section III, we compute the ground-state mass ranges of the $cc\bar{c}\bar{c}$ tetraquark for $J^P = 0^+$, $1^+$, and $2^+$. We also estimate the Regge slopes for the $0^+$, $1^+$, and $2^+$ trajectories, and determine the mass ranges for the orbitally excited states of the $cc\bar{c}\bar{c}$ and $ss\bar{c}\bar{c}$ tetraquarks in both the $(J, M^2)$ and $(n, M^2)$ planes. In Section IV, we discuss our results and Section V provides the conclusion of this work.

	\section{Theoretical Framework}
	The linear Regge trajectory is a popular phenomenological technique in hadron spectroscopy research. Almost every aspect of strong interactions is covered by regge theory, including particle spectra, forces between particles, and the behaviour of scattering amplitudes at high energies. A number of theories have been proposed to analyse the Regge trajectory. The simplest of them was Nambu's, which explained linear Regge trajectories and was given in 1970s \cite{ref40,ref41}.He assumed that a quark-antiquark pair interacts uniformly to form a strong flux tube, and that light quarks at the end of the tube rotate at the speed of light at radius $R$. The mass generated within this flux tube is calculated to be \cite{ref42}
	
	\begin{equation} \label{eq:1}
		M = 2 \int_{0}^{R} \frac{\sigma}{\sqrt{1 - \nu^2(r)}} \, dr = \pi \sigma R ,
	\end{equation}
	
	where $\sigma$ represents the string tension or mass density per unit length. Additionally, the angular momentum of the flux tube is computed as
	
	\begin{equation} \label{eq:2}
		J = 2 \int_{0}^{R} \frac{\sigma r \nu(r)}{\sqrt{1 - \nu^2(r)}} \, dr = \frac{\pi \sigma R^2}{2} + c'.
	\end{equation}
	Utilising equations (\ref{eq:1}) and (\ref{eq:2}), we can get the following formula.
	\begin{equation} \label{eq:3}
		J = \frac{M^2}{2\pi\sigma} + c'',
	\end{equation}
	where $c'$ and $c''$ are integration constants. Consequently, the relationship between $J$ and $M^2$ is linear. Chew-Frautschi plots are plots of hadron Regge trajectories in the $(J, M^2)$ plane \cite{ref43}. They utilized the theory to investigate the strong interaction between quarks and gluons. This study revealed that the experimentally absent higher excited states of mesons and baryons align with linear trajectories in the \((J, M^2)\) plane. \cite{ref43}. 
	
	Since both light and heavy hadrons exhibit quasilinear Regge trajectories, the most general expression for linear Regge trajectories can be written as follows \cite{ref39}:
	\begin{equation} \label{eq:4}
		J = \beta(M) = \beta(0) + \beta' M^2,
	\end{equation}
	where $\beta(0)$ represents the intercept and $\beta'$ denotes the slope of the particle's trajectory. Hadrons that share the same internal quantum numbers and lie on the same Regge trajectory are classified as part of the same family.
	
	From Eq.(\ref{eq:4}) we can have the following relation for the slope:
	
	\begin{equation} \label{eq:x}
		\beta'=\frac{(J+1)-J}{M^2_{(J+1)}-M^2_J}
	\end{equation}
	
	The Regge slopes and Regge intercepts for the various quark constituents of a meson multiplet with spin-parity $J^P$ (or more specifically, with quantum numbers $N^{2S+1} L_J$) can be related by the following expressions:
	
	\begin{equation} \label{eq:5}
		\beta_{i\bar{i}}(0) + \beta_{j\bar{j}}(0) = 2\beta_{i\bar{j}}(0),
	\end{equation}
	
	\begin{equation} \label{eq:6}
		\frac{1}{\beta'_{i\bar{i}}} + \frac{1}{\beta'_{j\bar{j}}} = \frac{2}{\beta'_{i\bar{j}}},
	\end{equation}
	where the quark flavors are represented by $i$ and $j$. A model utilizing the topological expansion and the quark-antiquark string representation of hadrons was employed to derive equations (\ref{eq:5}) and (\ref{eq:6}) \cite{ref47}. (Also see Refs.\cite{ref39,ref44,ref45,ref46}). The equation (\ref{eq:5}) was originally derived for light quarks in the dual resonance model \cite{ref48}. Later, it was found to hold true in the the dual-analytic model \cite{ref50}, two-dimensional QCD \cite{ref49}, and quark bremsstrahlung model \cite{ref51}.

	Here, and in the subsequent discussion, we focus on the case where the quark masses satisfy $m_i \leq m_j$ for two-body systems, as equations (\ref{eq:5}) and (\ref{eq:6}) remain symmetric under the exchange of quark flavors $i$ and $j$.
	
	\subsection{Relationship between slope ratios and masses}
	
	For two-body systems, solving equations (\ref{eq:5}) and (\ref{eq:6}) yields the following expression:
	\begin{equation} \label{eq:7}
		\beta'_{i\bar{i}} M_{i\bar{i}}^2 + \beta'_{j\bar{j}} M_{j\bar{j}}^2 = 2 \beta'_{i\bar{j}} M_{i\bar{j}}^2,
	\end{equation}
	By combining Eqs. (6) and (7), two sets of solutions are derived in terms of slope ratios and meson masses, expressed as:
	
\begin{equation} \label{eq:9}
	\frac{\beta'_{j\bar{j}}}{\beta'_{i\bar{i}}} = \frac{1}{2 M_{j\bar{j}}^2} \biggl[ \bigl(4 M_{i\bar{j}}^2 - M_{i\bar{i}}^2 - M_{j\bar{j}}^2 \bigr) \\
	\quad \pm \sqrt{\bigl(4 M_{i\bar{j}}^2 - M_{i\bar{i}}^2 - M_{j\bar{j}}^2 \bigr)^2 - 4 M_{i\bar{i}}^2 M_{j\bar{j}}^2} \biggr],
\end{equation}

and

\begin{equation} \label{eq:10}
	\frac{\beta'_{i\bar{j}}}{\beta'_{j\bar{j}}} = \frac{1}{4 M_{i\bar{j}}^2} \biggl[ \bigl(4 M_{i\bar{j}}^2 + M_{j\bar{j}}^2 - M_{i\bar{i}}^2 \bigr) \\
	\quad \pm \sqrt{\bigl(4 M_{i\bar{j}}^2 - M_{i\bar{i}}^2 - M_{j\bar{j}}^2 \bigr)^2 - 4 M_{i\bar{i}}^2 M_{j\bar{j}}^2} \biggr].
\end{equation}	
	

	In this work, we choose the solutions containing the plus sign preceding the square root term, as these yield slope ratios that closely match the experimentally observed slope ratios for certain well-known meson multiplets \cite{ref39}. Likewise, for tetraquark systems, when evaluating the slope ratios using Eq.~(\ref{eq:x}), the outcome aligns more closely with the result obtained from the solution with the plus sign, compared to that with the minus sign. This has been confirmed by comparing the ratio of $\beta'_{cc\bar{c}\bar{c}}$ to $\beta'_{bb\bar{b}\bar{b}}$, using theoretical mass values from Ref.~\cite{ref1050}.  
	Hence, both equations with the plus sign before the square root term can be expressed as:

	\begin{equation} \label{eq:100}
		\frac{\beta'_{j\bar{j}}}{\beta'_{i\bar{i}}} = \frac{1}{2 M_{j\bar{j}}^2} \biggl[ \bigl(4 M_{i\bar{j}}^2 - M_{i\bar{i}}^2 - M_{j\bar{j}}^2 \bigr) \\
		\quad + \sqrt{\bigl(4 M_{i\bar{j}}^2 - M_{i\bar{i}}^2 - M_{j\bar{j}}^2 \bigr)^2 - 4 M_{i\bar{i}}^2 M_{j\bar{j}}^2} \biggr],
	\end{equation}
	
	\begin{equation} \label{eq:200}
		\frac{\beta'_{i\bar{j}}}{\beta'_{j\bar{j}}} = \frac{1}{4 M_{i\bar{j}}^2} \biggl[ \bigl(4 M_{i\bar{j}}^2 + M_{j\bar{j}}^2 - M_{i\bar{i}}^2 \bigr) \\
		\quad + \sqrt{\bigl(4 M_{i\bar{j}}^2 - M_{i\bar{i}}^2 - M_{j\bar{j}}^2 \bigr)^2 - 4 M_{i\bar{i}}^2 M_{j\bar{j}}^2} \biggr].
	\end{equation}
	These equations provide significant relationships between the slope ratios and the masses of two-body systems.

	\subsection{Linear mass inequalities and quadratic mass inequalities}
	
	Equation~(\ref{eq:100}) leads to two important inequalities.
	
	Given that the Regge slopes $\alpha'_{j\bar{j}}$ and $\alpha'_{i\bar{i}}$ are required to be positive real quantities, their ratio $\alpha'_{j\bar{j}}/\alpha'_{i\bar{i}}$ must also be real. Therefore, from Eq.~(\ref{eq:100}), we derive

	\begin{equation} \label{eq:15}
		|4M^2_{i\bar{j}} - M^2_{i\bar{i}} - M^2_{j\bar{j}}| \geq 2M_{i\bar{i}}M_{j\bar{j}}.
	\end{equation}
	
	When \( i = j \), the condition \( 4M^2_{i\bar{j}} - M^2_{i\bar{i}} - M^2_{j\bar{j}} \leq 0 \) is not satisfied. Moreover, for \( i \neq j \), this inequality is inconsistent with the experimental data from well-established meson multiplets. Therefore, we infer that

	\begin{equation}
		4M^2_{i\bar{j}} - M^2_{i\bar{i}} - M^2_{j\bar{j}} \geq 0.
	\end{equation}  
	Consequently, Eq.~(\ref{eq:15}) can be reformulated as:
	
	\begin{equation} \label{eq:16}
		4M^2_{i\bar{j}} - M^2_{i\bar{i}} - M^2_{j\bar{j}} \geq 2M_{i\bar{i}}M_{j\bar{j}}.
	\end{equation}
	
	By adding \( M^2_{i\bar{i}} \) and \( M^2_{j\bar{j}} \) to both sides, we obtain:
	
	\begin{equation} \label{eq:17}
		2M_{i\bar{j}} \geq M_{i\bar{i}} + M_{j\bar{j}}.
	\end{equation}
	
	In the case where \( i = j \), it follows that \( M_{i\bar{i}} = M_{i\bar{j}} = M_{j\bar{j}} \), which implies the relation \( 2M_{i\bar{j}} = M_{i\bar{i}} + M_{j\bar{j}} \).
	
	On the other hand, even without assuming \( i = j \), if the condition \( 2M_{i\bar{j}} = M_{i\bar{i}} + M_{j\bar{j}} \) is satisfied, then Eq.~(\ref{eq:100}) allows us to derive the following:

	\begin{equation} \label{eq:18}
		\frac{\beta'_{j\bar{j}}}{\beta'_{i\bar{i}}} = \frac{M_{i\bar{i}}}{M_{j\bar{j}}}.
	\end{equation}
	
	The derivation of Eq.~(\ref{eq:18}) clearly shows that it is valid for mesons belonging to the same multiplet. Since hadrons lying on the same Regge trajectory possess the same slope, we arrive at

	\begin{equation} \label{eq:19}
		\frac{\beta'_{j\bar{j}}}{\beta'_{i\bar{i}}} = \frac{M_{i\bar{i},J}}{M_{j\bar{j},J}} = \frac{M_{i\bar{i},J+2}}{M_{j\bar{j},J+2}}.
	\end{equation}
	
	By applying Eq.~(\ref{eq:x}), one can calculate the slopes of particular Regge trajectories. For mesons composed of $i\bar{i}$ and $j\bar{j}$, the slopes are given by

	\begin{equation} \label{eq:20}
		\beta'_{i\bar{i}} = \frac{2}{M^2_{i\bar{i},J+2} - M^2_{i\bar{i},J}}, \quad \beta'_{j\bar{j}} = \frac{2}{M^2_{j\bar{j},J+2} - M^2_{j\bar{j},J}}. 
	\end{equation}
	
	Therefore using the above equation we can get,
	
	\begin{equation} \label{eq:21}
		\frac{\beta'_{j\bar{j}}}{\beta'_{i\bar{i}}} = \frac{M_{i\bar{i},J+2} + M_{i\bar{i},J}}{M_{j\bar{j},J+2} + M_{j\bar{j},J}} \times \frac{M_{i\bar{i},J+2} - M_{i\bar{i},J}}{M_{j\bar{j},J+2} - M_{j\bar{j},J}}. 
	\end{equation}
	
	Combining Eqs.~(\ref{eq:19}) and (\ref{eq:21}) yields

	\begin{equation} \label{eq:22}
		\frac{\beta'_{j\bar{j}}}{\beta'_{i\bar{i}}} = \frac{M_{i\bar{i},J+2} + M_{i\bar{i},J}}{M_{j\bar{j},J+2} + M_{j\bar{j},J}} \times \frac{M_{i\bar{i},J+2} - M_{i\bar{i},J}}{M_{j\bar{j},J+2} - M_{j\bar{j},J}} = \left( \frac{\beta'_{j\bar{j}}}{\beta'_{i\bar{i}}} \right)^2. 
	\end{equation}
	
	As mentioned earlier, the Regge slope $\beta'$ must be a positive real quantity. Therefore, based on Eq.~(\ref{eq:22}), the equality $\beta'_{j\bar{j}}/\beta'_{i\bar{i}} = 1$ is satisfied when $2M_{i\bar{j}} = M_{i\bar{i}} + M_{j\bar{j}}$. Consequently, Eq.~(\ref{eq:19}) leads to $M_{i\bar{i},J} = M_{j\bar{j},J}$ and $M_{i\bar{i},J+2} = M_{j\bar{j},J+2}$, indicating that $i = j$, assuming both $i\bar{i}$ and $j\bar{j}$ states have identical $J^P$.
	
	From the previous analysis, it can be concluded that the equation $2M_{i\bar{j}} = M_{i\bar{i}} + M_{j\bar{j}}$ is satisfied only when $i = j$. Therefore, for the case where $i \neq j$, Eq.~(\ref{eq:17}) gives

	\begin{equation} \label{eq:23}
		2M_{i\bar{j}} > M_{i\bar{i}} + M_{j\bar{j}}. 
	\end{equation}
	
	Using the above equation we get the following relation:
	\begin{equation} \label{eq:24}
		M_{j\bar{j}} < 2M_{i\bar{j}} - M_{i\bar{i}}. 
	\end{equation}
	
	Research has indicated that the slopes of Regge trajectories tend to decrease as the quark mass increases \cite{ref47,ref44,ref45,ref100,ref101,ref102,ref103,ref104,ref53}. Consequently, when the mass of the $j$ quark is larger than that of the $i$ quark, it follows that $\beta'_{j\bar{j}}/\beta'_{i\bar{i}} < 1$. Therefore, from Eq.~(\ref{eq:100}), one can derive

\begin{equation} \label{eq:25}
	\frac{1}{2M_{j\bar{j}}^{2}} \Bigg[ \left( 4M_{i\bar{j}}^{2} - M_{i\bar{i}}^{2} - M_{j\bar{j}}^{2} \right) \\
	+ \sqrt{\left( 4M_{i\bar{j}}^{2} - M_{i\bar{i}}^{2} - M_{j\bar{j}}^{2} \right)^{2} - 4M_{i\bar{i}}^{2} M_{j\bar{j}}^{2}} \Bigg] < 1
\end{equation}

	As the square root term in the above equation is positive, we can conclude that

	\begin{equation} \label{eq:26}
		2M_{j\bar{j}}^2 - (4M_{i\bar{j}}^2 - M_{i\bar{i}}^2 - M_{j\bar{j}}^2) > 0
	\end{equation}
	
	By Eqs. (\ref{eq:25}) and (\ref{eq:26}),	
	
\begin{equation} \label{eq:27}
	(4M_{i\bar{j}}^2 - M_{i\bar{i}}^2 - M_{j\bar{j}}^2)^2 
	- 4M_{i\bar{i}}^2 M_{j\bar{j}}^2 \\
	< \left[ 2M_{j\bar{j}}^2 - (4M_{i\bar{j}}^2 - M_{i\bar{i}}^2 - M_{j\bar{j}}^2) \right]^2
\end{equation}

	The last two equations can be used to get the following relation:
	\begin{equation} \label{eq:28}
		2M_{i\bar{j}}^2 < M_{i\bar{i}}^2 + M_{j\bar{j}}^2
	\end{equation}
	
	Therefore we get,
	\begin{equation} \label{eq:29}
		M_{j\bar{j}} > \sqrt{2M_{i\bar{j}}^2 - M_{i\bar{i}}^2}
	\end{equation}
	
	By applying Eqs.~(\ref{eq:24}) and (\ref{eq:29}), we obtain the following constraint relation for $M_{j\bar{j}}$.
	
	\begin{equation} \label{eq:30}
		\sqrt{2M_{i\bar{j}}^2 - M_{i\bar{i}}^2} < M_{j\bar{j}} < 2M_{i\bar{j}} - M_{i\bar{i}},
	\end{equation}
	
	The mass inequality outlined above establishes both the upper and lower limits for the mass of the $M_{j\bar{j}}$ meson. In the next section, we will use this relation to estimate the mass range of the tetraquarks that have yet to be discovered.

\section{Tetraquark Mass Spectra}
\subsection{The four-quark state in the diquark-antidiquark model}
In this study, we compute the mass spectra of all charm ($cc\bar{c}\bar{c}$) and doubly strange- doubly charm ($ss\bar{c}\bar{c}$) tetraquarks, treating them as bound states of two clusters (diquark and anti-diquark). The diquarks are considered as two coupled quarks, free from any internal spatial excitation. A diquark can only be observed within hadrons and treated as an effective degree of freedom because a pair of quarks cannot form a color singlet. A tetraquark in a color singlet state can be formed from two different diquark-antidiquark combinations: (i) a color anti-triplet diquark paired with a color triplet anti-diquark $\left( \overline{3} \otimes 3 \right)$, or (ii) a color sextet diquark paired with a color anti-sextet anti-diquark $\left( 6 \otimes \overline{6} \right)$.

If we treat the tetraquark as a two-body system composed of a diquark and an antidiquark, equation (\ref{eq:30}) can be used to determine the mass ranges of these states. Although this equation was originally derived under the assumption of linear Regge trajectories for light quark systems, several studies suggest that approximate linearity also extends to heavy–light and heavy–heavy systems \cite{Ebert:2011,Chen:2018,Ebert:2011:PRD}. While the Regge behavior of tetraquarks remains less well established, with some theoretical approaches predicting nonlinear or modified Regge trajectories for heavy hadrons \cite{Selem:2006,Cotugno:2009,Chen:2021}, phenomenological analyses often employ the linear approximation. This is because it provides reasonable fits to the known spectra of heavy–light and heavy–heavy systems. Therefore, despite theoretical complexities, we adopt the linear assumption in this work as a practical and widely supported approximation, and will use equation (\ref{eq:30}) to get mass ranges for tetraquark state.

\subsection{Mass Spectra of $cc\bar{c}\bar{c}$ and \(ss\bar{c}\bar{c}\) tetraquarks in the ($J,M^{2}$) plane}

Here, we use equation (\ref{eq:30}) to evaluate the mass range of ground-state of the $cc\bar{c}\bar{c}$ tetraquark. $cc\bar{c}\bar{c}$  tetraquark is considered to be composed of $cc$ diquark and $\bar{c}\bar{c}$ anti-diquark. Here, $c$ is charm quark. In eq (\ref{eq:30}), if we take $i=[ss]$, $j=[cc]$, we get the following relation. (Here, $s$ is the strange quark.)

\begin{equation} \label{eq:31}
	\sqrt{2M_{ss\bar{c}\bar{c}}^2 - M_{ss\bar{s}\bar{s}}^2} < M_{cc\bar{c}\bar{c}} < 2M_{ss\bar{c}\bar{c}} - M_{ss\bar{s}\bar{s}},
\end{equation}

In this study, we take the masses of the $ss\bar{c}\bar{c}$ and $ss\bar{s}\bar{s}$ states from Refs.~\cite{Braaten:2021} and \cite{ref55}, respectively, as theoretical inputs due to the absence of experimental data. By inserting the theoretical masses of $ss\bar{s}\bar{s}$ and $ss\bar{c}\bar{c}$ with quantum numbers $J^P = 0^+$, $1^+$, and $2^+$ into our framework, we determine the ground-state mass ranges for the $cc\bar{c}\bar{c}$ tetraquark as 5.712--6.411 GeV for $0^+$, 5.733--6.425 GeV for $1^+$, and 5.778--6.458 GeV for $2^+$.

To estimate the higher excited states, we compute the Regge slopes corresponding to these tetraquark systems. Specifically, to evaluate the value of $\beta'$ for the $cc\bar{c}\bar{c}$ configuration, we utilize Eq.~(\ref{eq:100}). By inserting suitable values for $i$ and $j$ and solving for $\beta'_{cc\bar{c}\bar{c}}$, we derive the following expression:

\begin{equation} \label{eq:32}
		\beta'_{cc\bar{c}\bar{c}} = \frac{\beta'_{ss\bar{s}\bar{s}}}{2 M_{cc\bar{c}\bar{c}}^2} \biggl[ \bigl(4 M_{ss\bar{c}\bar{c}}^2 - M_{ss\bar{s}\bar{s}}^2 - M_{cc\bar{c}\bar{c}}^2 \bigr) \\
		\quad + \sqrt{\bigl(4 M_{ss\bar{c}\bar{c}}^2 - M_{ss\bar{s}\bar{s}}^2 - M_{cc\bar{c}\bar{c}}^2 \bigr)^2 - 4 M_{ss\bar{s}\bar{s}}^2 M_{cc\bar{c}\bar{c}}^2} \biggr]
\end{equation}

We can find the slope of the Regge trajectory for \(ss\bar{s}\bar{s}\) tetraquark using equation (\ref{eq:x}),

\begin{equation} \label{eq:33}
	\beta'_{ss\bar{s}\bar{s}} = \frac{1}{M_{ss\bar{s}\bar{s}(1^-)}^2 - M_{ss\bar{s}\bar{s}(0^+)}^2} .
\end{equation} 

The mass of the \(ss\bar{s}\bar{s}\) tetraquark with \(J^P = 1^-\), taken from Ref.~\cite{ref55}, is used to compute its Regge slope. The resulting slope values for different \(J^P\) quantum numbers are listed in Table~\ref{tab:Slope}.

By inserting the values of \(M_{ss\bar{s}\bar{s}}\), \(M_{ss\bar{c}\bar{c}}\), and \(\beta'_{ss\bar{s}\bar{s}}\) into Eq.~(\ref{eq:31}), \(\beta'_{cc\bar{c}\bar{c}}\) can be represented as a function of \(M_{cc\bar{c}\bar{c}}\). This function decreases over the interval (5.712--6.411). For \(J^P = 0^+\), the estimated range of \(\beta'_{cc\bar{c}\bar{c}}\) is from 0.21156 to 0.59149, as shown in Table~\ref{tab:Slope}. The corresponding ranges for other \(J^P\) values are also included in the same table.

By using equation (\ref{eq:6}) and (\ref{eq:32}) we can get the below equation:
\\
\\

	\begin{equation} \label{eq:101}
		\beta'_{ss\bar{c}\bar{c}} = \frac{2}{
			\left(
			\frac{1}{\beta'_{ss\bar{s}\bar{s}}} + \frac{1}{\beta'_{ss\bar{s}\bar{s}} \cdot \frac{1}{2 M_{cc\bar{c}\bar{c}}^{2}} \left( \left(4 M_{ss\bar{c}\bar{c}}^{2} - M_{ss\bar{s}\bar{s}}^{2} - M_{cc\bar{c}\bar{c}}^{2} \right) + \sqrt{ \left(4 M_{ss\bar{c}\bar{c}}^{2} - M_{ss\bar{s}\bar{s}}^{2} - M_{cc\bar{c}\bar{c}}^{2} \right)^{2} - 4 M_{ss\bar{s}\bar{s}}^{2} M_{cc\bar{c}\bar{c}}^{2} } \right)}
			\right)
		}
	\end{equation}

By plugging the values of \(M_{ss\bar{s}\bar{s}}\), \(M_{ss\bar{c}\bar{c}}\), and \(\beta'_{ss\bar{s}\bar{s}}\) into the equation above, \(\beta'_{ss\bar{c}\bar{c}}\) can be formulated as a function of \(M_{cc\bar{c}\bar{c}}\). Within the interval 5.712--6.411, this function exhibits a decreasing trend. For \(J^P = 0^+\), the corresponding range of \(\beta'_{ss\bar{c}\bar{c}}\) is from 0.31165 to 0.59149, as listed in Table~\ref{tab:Slope}. The ranges for other \(J^P\) states are also presented in the same table.

Furthermore, using Eq.~(\ref{eq:x}), the mass of the excited $cc\bar{c}\bar{c}$ tetraquark state can be expressed as:
\begin{equation} \label{eq:34}
	M_{J+k(cc\bar{c}\bar{c})} = \sqrt{M_{J(cc\bar{c}\bar{c})}^2 + \frac{k}{\beta'_{cc\bar{c}\bar{c}}}} ,
\end{equation}
where, k is an positive integer number.

Using Eqs.~(\ref{eq:34}) and (\ref{eq:32}), we can express the mass \( M_{J+k(cc\bar{c}\bar{c})} \) in terms of \( M_{J(cc\bar{c}\bar{c})} \).  
For example, when \( J^P = 0^+ \) and \( k = 1 \), the function is increasing over the range (5.712-6.411). The mass for \( cc\bar{c}\bar{c} \) with \( J^P = 1^- \) lies between 5.858 GeV and 6.770 GeV, as shown in Table~\ref{table:all charm_tetraquarks}. Likewise, the masses of other excited states for the \( cc\bar{c}\bar{c} \) tetraquark are calculated and presented in Table~\ref{table:all charm_tetraquarks}.

\begin{table}[h]
	\caption{Values of Regge Slopes for \(ss\bar{s}\bar{s}\), \(cc\bar{c}\bar{c}\) and \(ss\bar{c}\bar{c}\) tetraquarks in \((J,M^2)\) plane (in \(\text{GeV}^{-2}\))}{\label{tab:Slope}}
	\begin{tabular}{cccc}
		\toprule
		\(J^P\) & \(\beta'_{ss\bar{s}\bar{s}} \, (\text{GeV}^{-2})\) & \(\beta'_{cc\bar{c}\bar{c}} \, (\text{GeV}^{-2})\) & \(\beta'_{ss\bar{c}\bar{c}} \, (\text{GeV}^{-2})\) \\
		\midrule
		\(0^+\) & 0.59149 & 0.21156-0.59149 & 0.31165-0.59149 \\
		\(1^+\) & 0.57189 & 0.20677-0.57189 & 0.30372-0.57189 \\
		\(2^+\) & 0.57535 & 0.21186-0.57535 & 0.30968-0.57535 \\
		\botrule
	\end{tabular}
\end{table}

In a same way, we can get the corresponding formula for the \(ss\bar{c}\bar{c}\) tetraquark using Eq. (\ref{eq:x}).

\begin{equation} \label{eq:35}
	M_{J+k(ss\bar{c}\bar{c})} = \sqrt{M_{J(ss\bar{c}\bar{c})}^2 + \frac{k}{\beta'_{ss\bar{c}\bar{c}}}} ,
\end{equation}

By using Eqs.~(\ref{eq:35}) and (\ref{eq:101}), we can express \( M_{J+k(ss\bar{c}\bar{c})} \) as a function of \( M_{J(cc\bar{c}\bar{c})} \). With this method, we have computed the mass ranges for the excited states of the \( ss\bar{c}\bar{c} \) tetraquark, which are summarized in Table~\ref{table:sscc_tetraquarks}.

The estimated mass spectra for the \(cc\bar{c}\bar{c}\) and \(ss\bar{c}\bar{c}\) systems are shown in Tables~\ref{table:all charm_tetraquarks} and~\ref{table:sscc_tetraquarks}, respectively, as mentioned earlier. In addition, we have compared our results with the two-meson threshold values. Furthermore, the results for the all-charm tetraquark are also compared with those from other studies in Table~\ref{table:all_charm_tetraquarks_comparison}. And Table~\ref{table:sscc_tetraquarks} also contains comparison of our calculated values for $ss\bar{c}\bar{c}$ with opther prediction.

\begin{figure}[htbp]
	\centering
	\includegraphics[width=\linewidth]{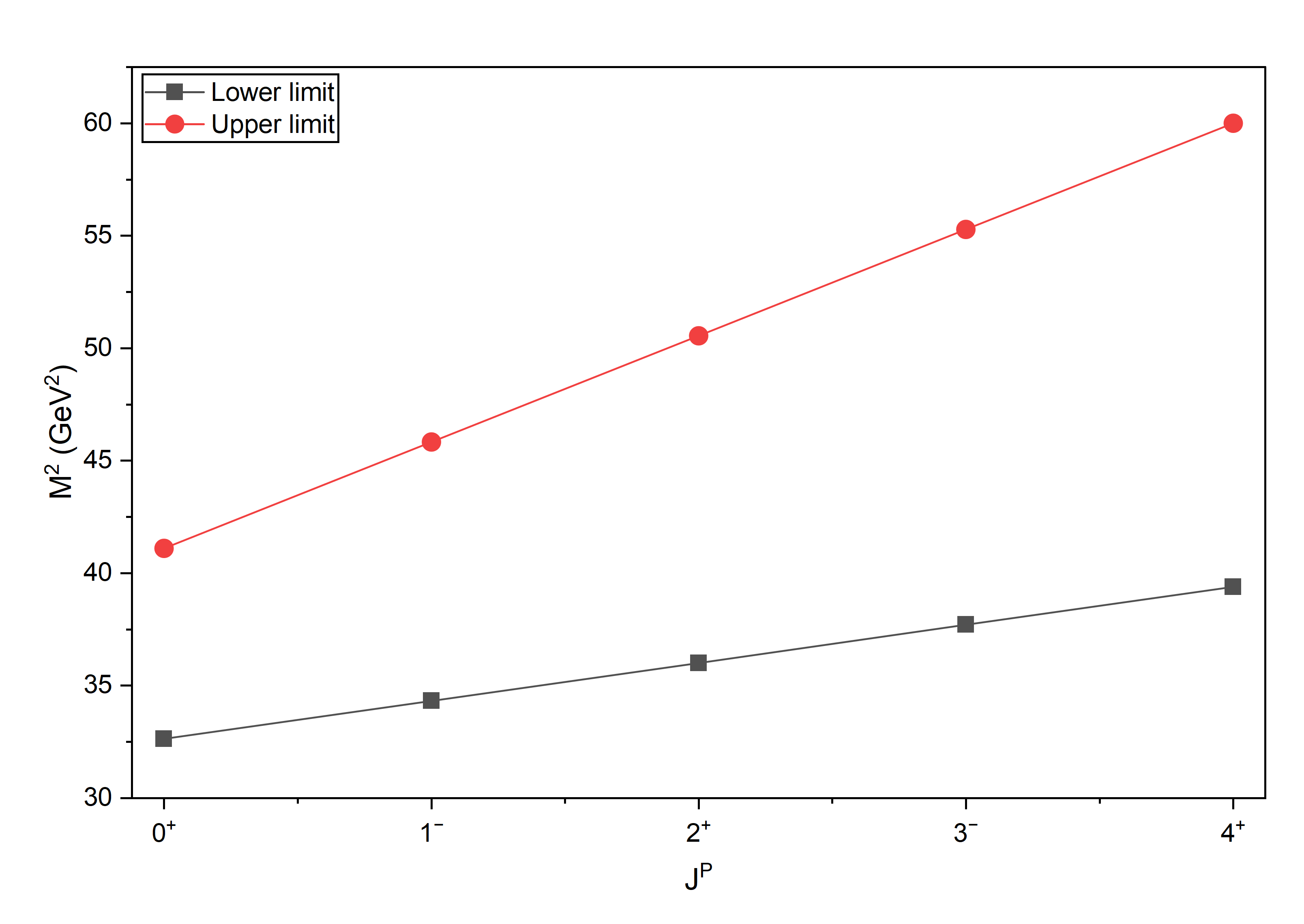}  
	\caption{Regge trajectory of $cc\bar{c}\bar{c}$ tetraqaurak for $S=0$ in $(J,M^2)$ plane.}
	\label{fig:regge trajectory for cccc of S=0 in J plane}
\end{figure}

\begin{figure}[htbp]
	\centering
	\includegraphics[width=\linewidth]{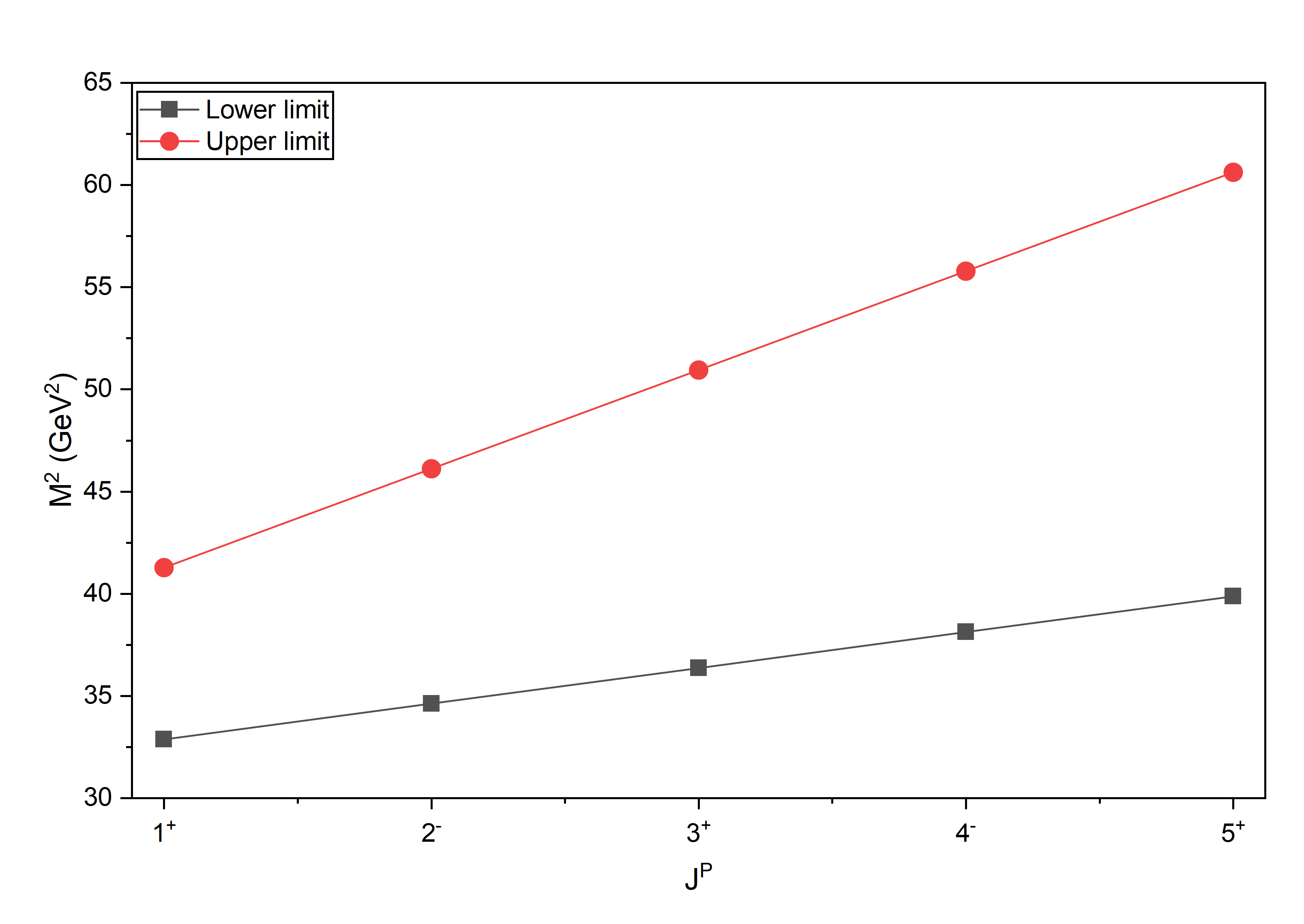}  
	\caption{Regge trajectory of $cc\bar{c}\bar{c}$ tetraqaurak for $S=1$ in $(J,M^2)$ plane.}
	\label{fig:regge trajectory for cccc of S=1 in J plane}
\end{figure}

\begin{figure}[htbp]
	\centering
	\includegraphics[width=\linewidth]{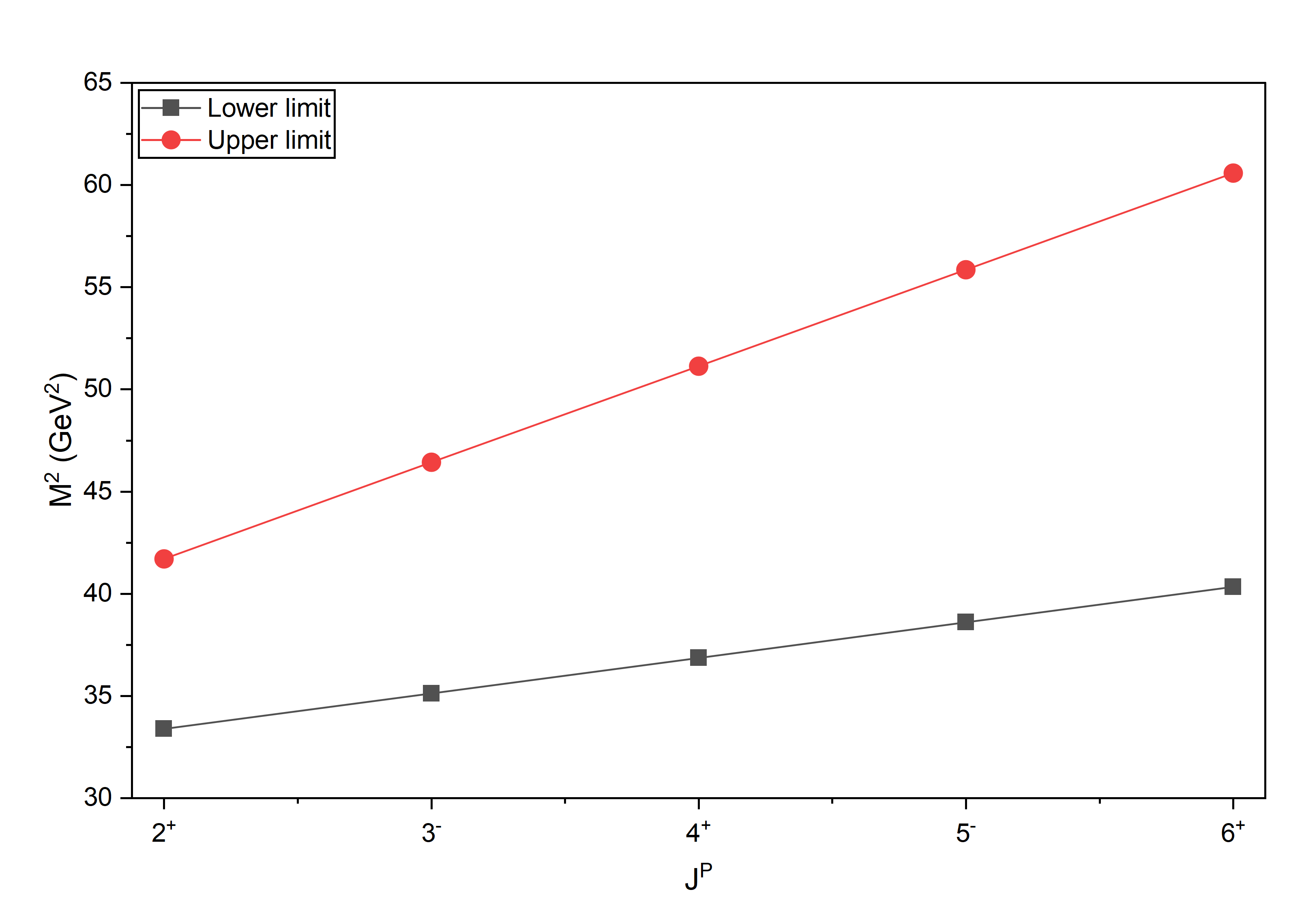}  
	\caption{Regge trajectory of $cc\bar{c}\bar{c}$ tetraqaurak for $S=2$ in $(J,M^2)$ plane.}
	\label{fig:regge trajectory for cccc of S=2 in J plane}
\end{figure}

\begin{figure}[htbp]
	\centering
	\includegraphics[width=\linewidth]{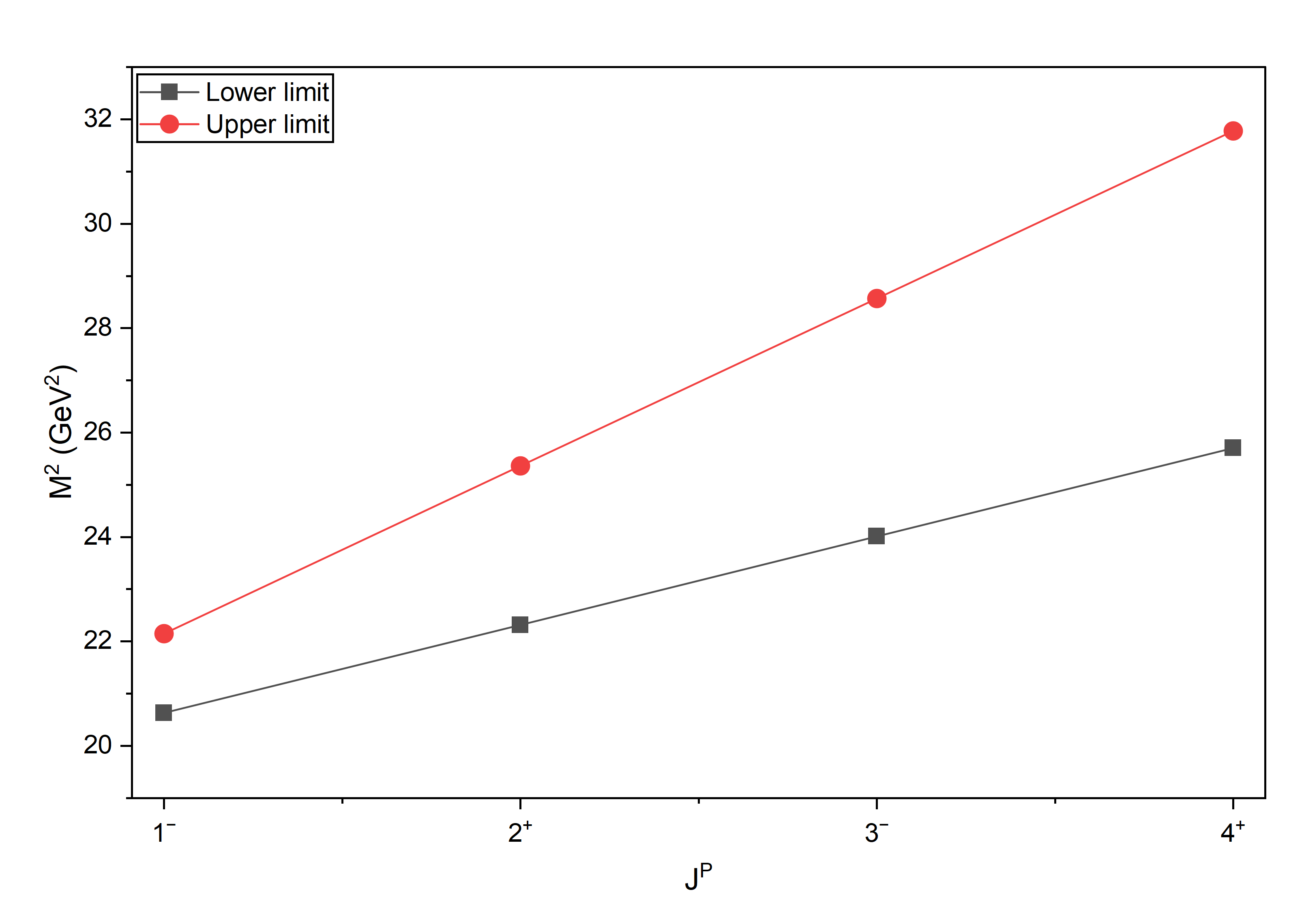}  
	\caption{Regge trajectory of $ss\bar{c}\bar{c}$ tetraqaurak for $S=0$ in $(J,M^2)$ plane.}
	\label{fig:regge trajectory for sscc of S=0 in J plane}
\end{figure}

\begin{figure}[htbp]
	\centering
	\includegraphics[width=\linewidth]{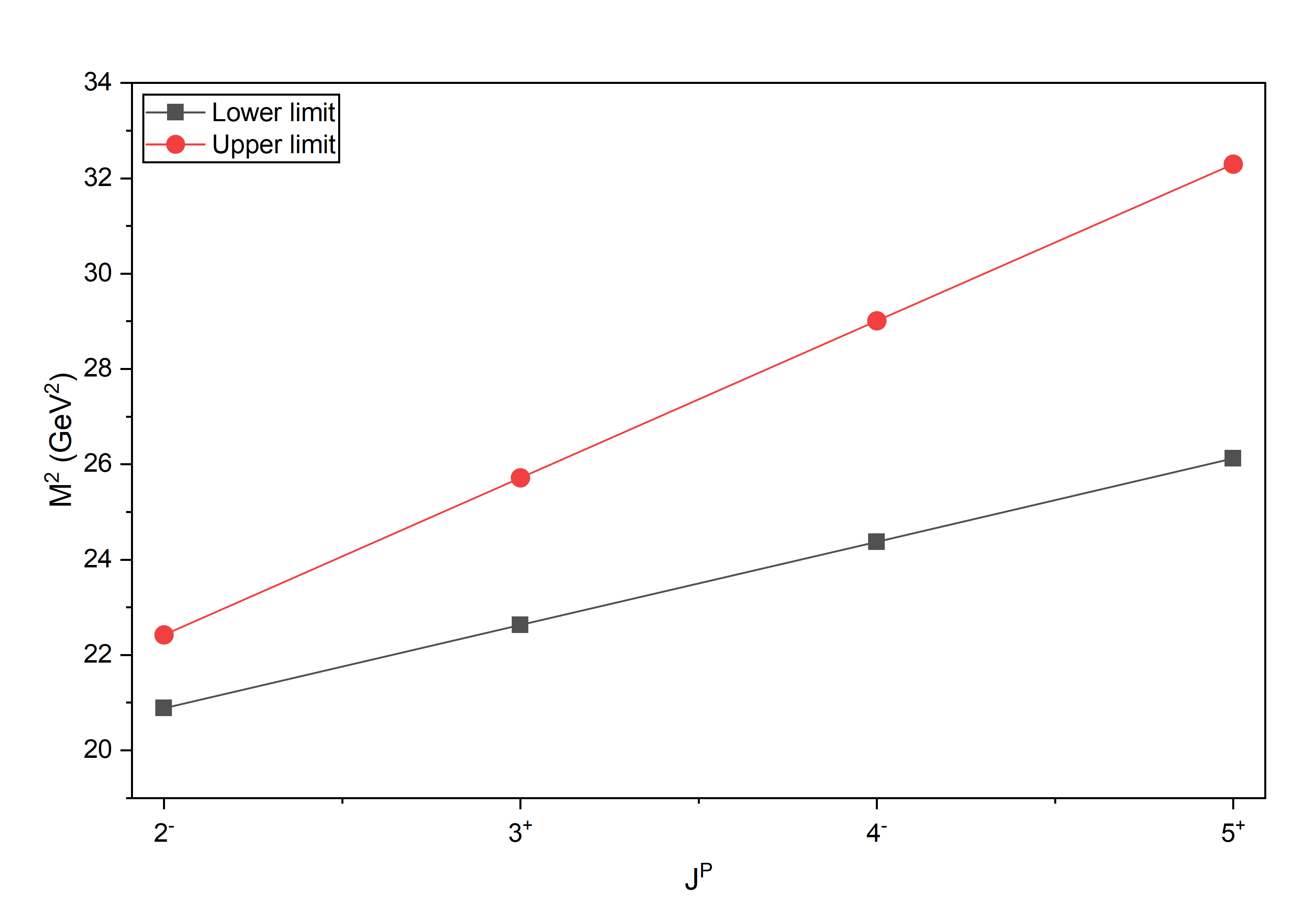}  
	\caption{Regge trajectory of $ss\bar{c}\bar{c}$ tetraqaurak for $S=1$ in $(J,M^2)$ plane.}
	\label{fig:regge trajectory for sscc of S=1 in J plane}
\end{figure}

\begin{figure}[htbp]
	\centering
	\includegraphics[width=\linewidth]{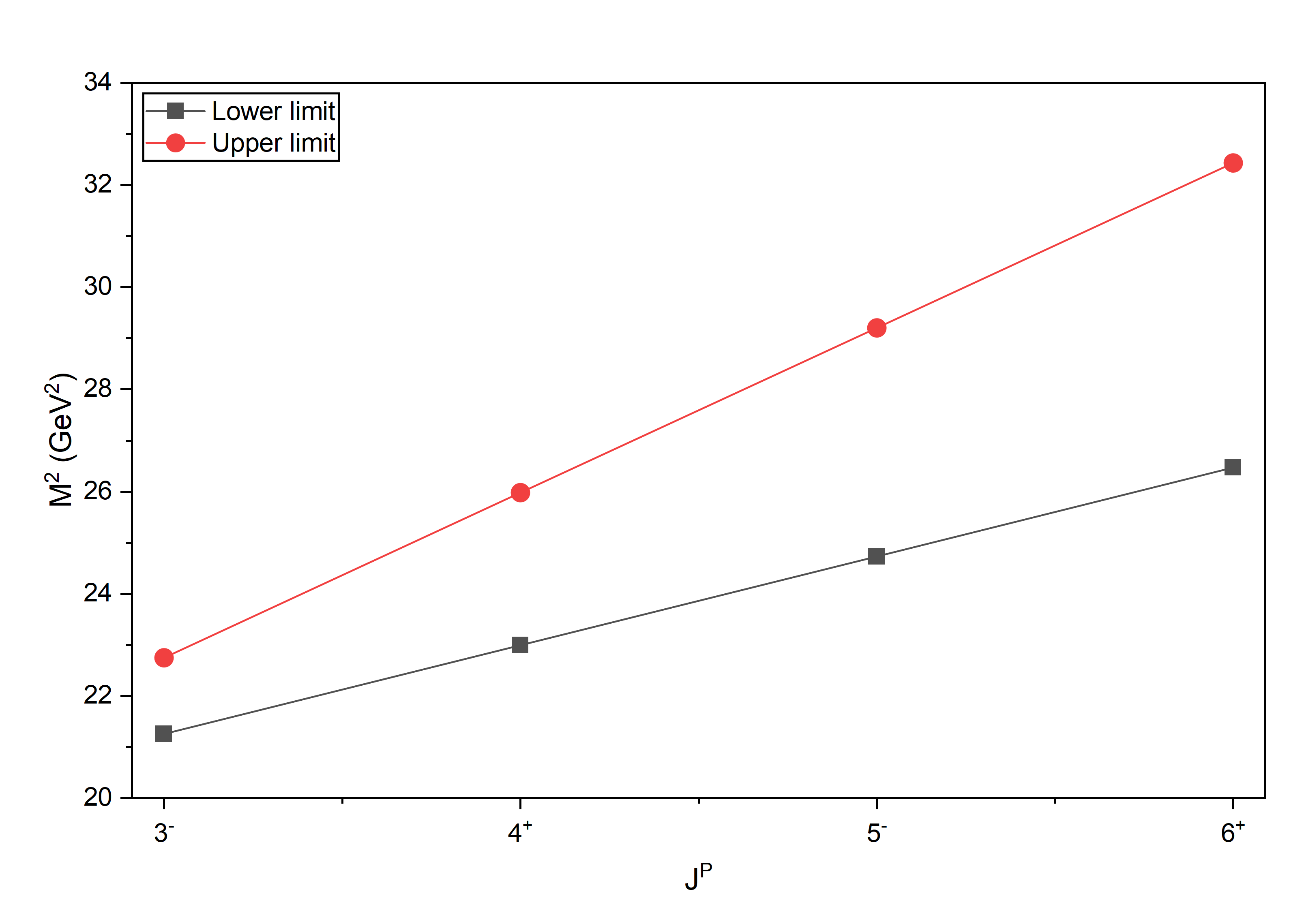}  
	\caption{Regge trajectory of $ss\bar{c}\bar{c}$ tetraqaurak for $S=2$ in $(J,M^2)$ plane.}
	\label{fig:regge trajectory for sscc of S=2 in J plane}
\end{figure}

Furthermore, we have plotted the Regge trajectories for the $cc\bar{c}\bar{c}$ tetraquark in the $(J,M^2)$ plane, as presented in Figures~\ref{fig:regge trajectory for cccc of S=0 in J plane}, \ref{fig:regge trajectory for cccc of S=1 in J plane}, and \ref{fig:regge trajectory for cccc of S=2 in J plane} corresponding to different spin values. Each of these plots displays two trajectories: one representing the lower mass limits and the other representing the upper mass limits. In a similar manner, the trajectories for the $ss\bar{c}\bar{c}$ tetraquark in the $(J,M^2)$ plane are shown in Figures~\ref{fig:regge trajectory for sscc of S=0 in J plane}, \ref{fig:regge trajectory for sscc of S=1 in J plane}, and \ref{fig:regge trajectory for sscc of S=2 in J plane}.

\begin{table}[htb]
	\centering
	\caption{Mass spectra of all charm tetraquarks with their corresponding two-meson thresholds.}
	\label{table:all charm_tetraquarks}
	\begin{tabular}{ccccc}
		\toprule
		State & $J^P$ & Calculated & Two-meson & Threshold \\ 
		&       & Mass & threshold & Mass  \\ 
		&       & (GeV) &  & (GeV) \\ 
		\midrule
		$1^1S_0$ & $0^+$ & 5.712-6.411 & $\eta_c(1S)\eta_c(1S)$ & 5.968 \\
		$1^1P_1$ & $1^-$ & 5.858-6.770 & $\eta_c(1S)\chi_{c1}(1P)$ & 6.495 \\
		$1^1D_2$ & $2^+$ & 6.000-7.110 & $J/\psi(1S)J/\psi(1S)$ & 6.194 \\
		$1^1F_3$ & $3^-$ & 6.140-7.435 & $J/\psi(1S)\chi_{c2}(1P)$ & 6.653 \\
		$1^1G_4$ & $4^+$ & 6.276-7.746 & $J/\psi(1S)\psi_3(3842)$ & 6.940 \\
		$1^3S_1$ & $1^+$ & 5.733-6.425 & $\eta_c(1S)J/\psi(1S)$ & 6.081 \\
		$1^3P_2$ & $2^-$ & 5.884-6.791 & $\eta_c(1S)\chi_{c2}(1P)$ & 6.540 \\
		$1^3D_3$ & $3^+$ & 6.030-7.138 & $\eta_{c}(1S)\psi_3(3842)$ & 6.827 \\
		$1^3F_4$ & $4^-$ & 6.174-7.469 & $\chi_{c1}(1P)\psi_3(3842)$ & 7.354 \\
		$1^3G_5$ & $5^+$ & 6.314-7.786 & $\psi_2(3823)\psi_3(3842)$ & 7.667 \\
		$1^5S_2$ & $2^+$ & 5.778-6.458 & $J/\psi(1S)J/\psi(1S)$ & 6.194 \\
		$1^5P_3$ & $3^-$ & 5.926-6.814 & $\psi_{c2}(1P)J/\psi(1S)$ & 6.653 \\
		$1^5D_4$ & $4^+$ & 6.071-7.151 & $J/\psi(1S)\psi_3(3842)$ & 6.940 \\
		$1^5F_5$ & $5^-$ & 6.213-7.474 & $\psi_3(3842)\chi_{c2}(1P)$ & 7.399 \\
		$1^5G_6$ & $6^+$ & 6.351-7.784 & $\psi_3(3842)\psi_3(3842)$ & 7.686 \\
		\botrule
	\end{tabular}
\end{table}

\begin{table*}[t]
	\centering
	\scriptsize
	\setlength{\tabcolsep}{1.8pt}
	\caption{Comparison of all-charm tetraquark ($cc\bar{c}\bar{c}$) masses with other studies (in GeV).}
	\label{table:all_charm_tetraquarks_comparison}
	\resizebox{\textwidth}{!}{%
		\begin{tabular}{c c c c c c c c c c c c c c c c c c c}
			\toprule
			State & $J^P$ & Ours & \cite{Tiwari1:2023} & \cite{Wang:2017} &
			\cite{Faustov:2021} & \cite{Lundhammar:2020} & \cite{Debastiani:2019} &
			\cite{Bedolla:2020} & \cite{Weng:2021} & \cite{Lloyd:2004} &
			\cite{Gordillo:2020} & \cite{Berezhnoy:2012} & \cite{Ader:1982} &
			\cite{Liu:2019} & \cite{Zhao:2020} & \cite{Zhao:2021} &
			\cite{RZhu:2021} & \cite{Chen:2017} \\
			\midrule
			$1^1 S_0$ & $0^+$ & 5.712--6.411 & 5.942 & 5.990 & 6.190 & 5.960 & 5.969 & 5.883 & 6.044 & 5.960 & 6.351 & 5.966 & 6.437 & 6.518 & 6.346 & 6.466 & 6.055 & 6.440 \\
			$1^1 P_1$ & $1^-$ & 5.858--6.770 & 6.555 &  & 6.631 &  & 6.577 &  &  &  &  & 6.718 &  &  &  &  &  & 6.830 \\
			$1^1 D_2$ & $2^+$ & 6.000--7.110 &  &  & 6.921 &  &  &  &  &  &  &  &  &  &  &  &  &  \\
			$1^1 F_3$ & $3^-$ & 6.140--7.435 &  &  &  &  &  &  &  &  &  &  &  &  &  &  &  &  \\
			$1^1 G_4$ & $4^+$ & 6.276--7.746 &  &  &  &  &  &  &  &  &  &  &  &  &  &  &  &  \\
			$1^3 S_1$ & $1^+$ & 5.733--6.425 & 5.989 &  & 6.271 & 6.009 & 6.021 & 6.120 & 6.230 & 6.009 & 6.441 & 6.051 &  & 6.500 & 6.441 & 6.494 &  & 6.370 \\
			$1^3 P_2$ & $2^-$ & 5.884--6.791 & 6.589 &  & 6.644 &  & 6.609 &  &  &  &  &  &  &  &  &  &  &  \\
			$1^3 D_3$ & $3^+$ & 6.030--7.138 &  &  & 6.932 &  &  &  &  &  &  &  &  &  &  &  &  &  \\
			$1^3 F_4$ & $4^-$ & 6.174--7.469 &  &  &  &  &  &  &  &  &  &  &  &  &  &  &  &  \\
			$1^3 G_5$ & $5^+$ & 6.314--7.786 &  &  &  &  &  &  &  &  &  &  &  &  &  &  &  &  \\
			$1^5 S_2$ & $2^+$ & 5.778--6.458 & 6.082 & 6.090 & 6.367 & 6.100 & 6.115 & 6.246 & 6.287 & 6.100 & 6.471 & 6.223 &  & 6.524 & 6.475 & 6.551 & 6.090 & 6.510 \\
			$1^5 P_3$ & $3^-$ & 5.926--6.814 & 6.625 &  & 6.664 &  & 6.641 &  &  &  &  &  &  &  &  &  &  &  \\
			$1^5 D_4$ & $4^+$ & 6.071--7.151 &  &  & 6.945 &  &  &  &  &  &  &  &  &  &  &  &  &  \\
			$1^5 F_5$ & $5^-$ & 6.213--7.474 &  &  &  &  &  &  &  &  &  &  &  &  &  &  &  &  \\
			$1^5 G_6$ & $6^+$ & 6.351--7.784 &  &  &  &  &  &  &  &  &  &  &  &  &  &  &  &  \\
			\botrule
		\end{tabular}%
	}
\end{table*}

\begin{table}[htb]
	\centering
	\caption{Mass spectra of $ss\bar{c}\bar{c}$ tetraquarks with their corresponding two-meson thresholds and comparison with \cite{Tiwari:2023}}
	\label{table:sscc_tetraquarks}
	\begin{tabular}{cccccc}
		\toprule
		State & $J^P$ & Calculated & Two-meson & Threshold & Ref. \cite{Tiwari:2023}\\ 
		&       & Mass & threshold & Mass &   \\ 
		&       & (GeV) &  & (GeV) &  \\ 
		\midrule
		$1^1P_1$ & $1^-$ & 4.542-4.706 & $D^{*\pm}_sD^{*}_{s0}(2317)^\pm$ & 4.430 & 4.556  \\
		$1^1D_2$ & $2^+$ & 4.724-5.036 & $D^{*\pm}_sD^{*\pm}_s$ & 4.224 & \\
		$1^1F_3$ & $3^-$ & 4.900-5.345 & $D^{*\pm}_sD^{*}_{s2}(2573)$ & 4.681 & \\
		$1^1G_4$ & $4^+$ & 5.070-5.637 & $D^{*\pm}_sD^{*}_{s3}(2860)^\pm$ & 4.912 & \\
		$1^3P_2$ & $2^-$ & 4.570-4.735 & $D^{*\pm}_sD_{s1}(2460)^\pm$ & 4.572 & 4.581 \\
		$1^3D_3$ & $3^+$ & 4.757-5.071 & $D^{*\pm}_sD^{*}_{s2}(2573)$ & 4.681 & \\
		$1^3F_4$ & $4^-$ & 4.937-5.386 & $D_{s1}(2460)^{\pm}D^{*}_{s3}(2860)^\pm$ & 5.320 & \\
		$1^3G_5$ & $5^+$ & 5.111-5.683 & -- & -- & \\
		$1^5P_3$ & $3^-$ & 4.610-4.769 & $D^{*\pm}_sD^{*}_{s2}(2573)$ & 4.681 & 4.612 \\
		$1^5D_4$ & $4^+$ & 4.795-5.097 & $D^{*\pm}_sD^{*}_{s3}(2860)^\pm$ & 4.972 & \\
		$1^5F_5$ & $5^-$ & 4.973-5.404 & $D^{*}_{s2}(2573)D^{*}_{s3}(2860)^\pm$ & 5.429 & \\
		$1^5G_6$ & $6^+$ & 5.145-5.695 & $D^{*}_{s3}(2860)^{\pm}D^{*}_{s3}(2860)^\pm$ & 5.720 & \\
		\botrule
	\end{tabular}
\end{table}

\subsection{Mass Spectra of $cc\bar{c}\bar{c}$ and \(ss\bar{c}\bar{c}\) tetraquarks in the $(n,M^2)$ plane}

In this section, we will calculate the Regge parameters for the \(cc\bar{c}\bar{c}\) and \(ss\bar{c}\bar{c}\) tetraquarks in the \((n, M^2)\) plane to assess the masses of the radial excited states. The general form of the linear equation for the Regge trajectory in the \((n, M^2)\) plane is given by:

\begin{equation} \label{eq:36}
	n = \alpha(M) = \alpha(0) + \alpha' M^2 ,
\end{equation}

where \(n = 1, 2, 3, \ldots\) denotes the radial principal quantum number, and \(\alpha(0)\) and \(\alpha'\) represent the intercept and slope of the trajectory in the \((n, M^2)\) plane. It is assumed that the Regge parameters are the same for all tetraquark multiplets situated along the same Regge line. In this case, we have employed a similar approach to determine the Regge parameters as we did previously in the \((J, M^2)\) plane.

From Eq. (\ref{eq:36}), we can determine the slope for $ss\bar{s}\bar{s}$ tetraquark in $(n,M^2)$ plane by the following equation 
\begin{equation} \label{eq:37}
	\alpha'_{ss\bar{s}\bar{s}} = \frac{1}{M_{ss\bar{s}\bar{s}(2S)}^2 - M_{ss\bar{s}\bar{s}(1S)}^2}, 
\end{equation}

Due to the lack of experimental data, we have used the theoretically predicted masses of the \( 1^1S_0 \) and \( 2^1S_0 \) states of the all-strange (\( ss\bar{s}\bar{s} \)) tetraquark from Ref.~\cite{ref55} for our analysis. Utilizing Eq.~(\ref{eq:37}), we calculate the Regge slope \( \alpha'_{ss\bar{s}\bar{s}} = 0.35544 \, \text{GeV}^{-2} \), which is reported in Table~\ref{tab:SlopenM2} along with slopes corresponding to other \( J^P \) states.

It is assumed that Eqs.~(\ref{eq:5}) and (\ref{eq:6}), which are valid in the \( (J, M^2) \) plane, are equally applicable in the \( (n, M^2) \) plane. Therefore, using Eq.~(\ref{eq:100}), we derive the following expression in the \( (n, M^2) \) framework:

\begin{equation} \label{eq:38}
		\alpha'_{cc\bar{c}\bar{c}} = \frac{\alpha'_{ss\bar{s}\bar{s}}}{2 M_{cc\bar{c}\bar{c}}^2} \Biggl[ \Bigl(4 M_{ss\bar{c}\bar{c}}^2 - M_{ss\bar{s}\bar{s}}^2 - M_{cc\bar{c}\bar{c}}^2 \Bigr) \\
		\quad + \sqrt{\Bigl(4 M_{ss\bar{c}\bar{c}}^2 - M_{ss\bar{s}\bar{s}}^2 - M_{ss\bar{c}\bar{c}}^2 \Bigr)^2 - 4 M_{ss\bar{s}\bar{s}}^2 M_{cc\bar{c}\bar{c}}^2} \Biggr].
\end{equation}	

By inserting the ground-state (\(1^1 S_0\)) masses of the \(ss\bar{s}\bar{s}\) and \(ss\bar{c}\bar{c}\) tetraquarks from Refs.~\cite{ref55} and ~\cite{Braaten:2021}, respectively, along with the slope value \(\alpha'_{ss\bar{s}\bar{s}} = 0.35544\), into the above relation, \(\alpha'_{cc\bar{c}\bar{c}}\) can be formulated as a function of \(M_{cc\bar{c}\bar{c}}\). This function shows a decreasing behavior within the interval (5.412--6.411), and the resulting range for \(\alpha'_{cc\bar{c}\bar{c}}\) is from 0.12713 to 0.35544, as summarized in Table~\ref{tab:SlopenM2}.

Using a method analogous to that applied in the \((J, M^2)\) plane, the slope parameters for other tetraquark systems, such as the \(ss\bar{c}\bar{c}\) state, have been determined in the \((n, M^2)\) plane. All slope values derived in this framework are summarized in Table~\ref{tab:SlopenM2}. Furthermore, employing the same strategy as used for computing the excited-state masses in the \((J, M^2)\) plane, we have calculated the excited-state mass spectra of the \(cc\bar{c}\bar{c}\) and \(ss\bar{c}\bar{c}\) tetraquarks in the \((n, M^2)\) plane. The results are presented in Tables~\ref{tab:cccc_spectra_innm^2} and \ref{tab:sscc_spectra_innm^2}, respectively, and are compared with existing theoretical predictions.

\begin{figure}[htbp]
	\centering
	\includegraphics[width=\linewidth]{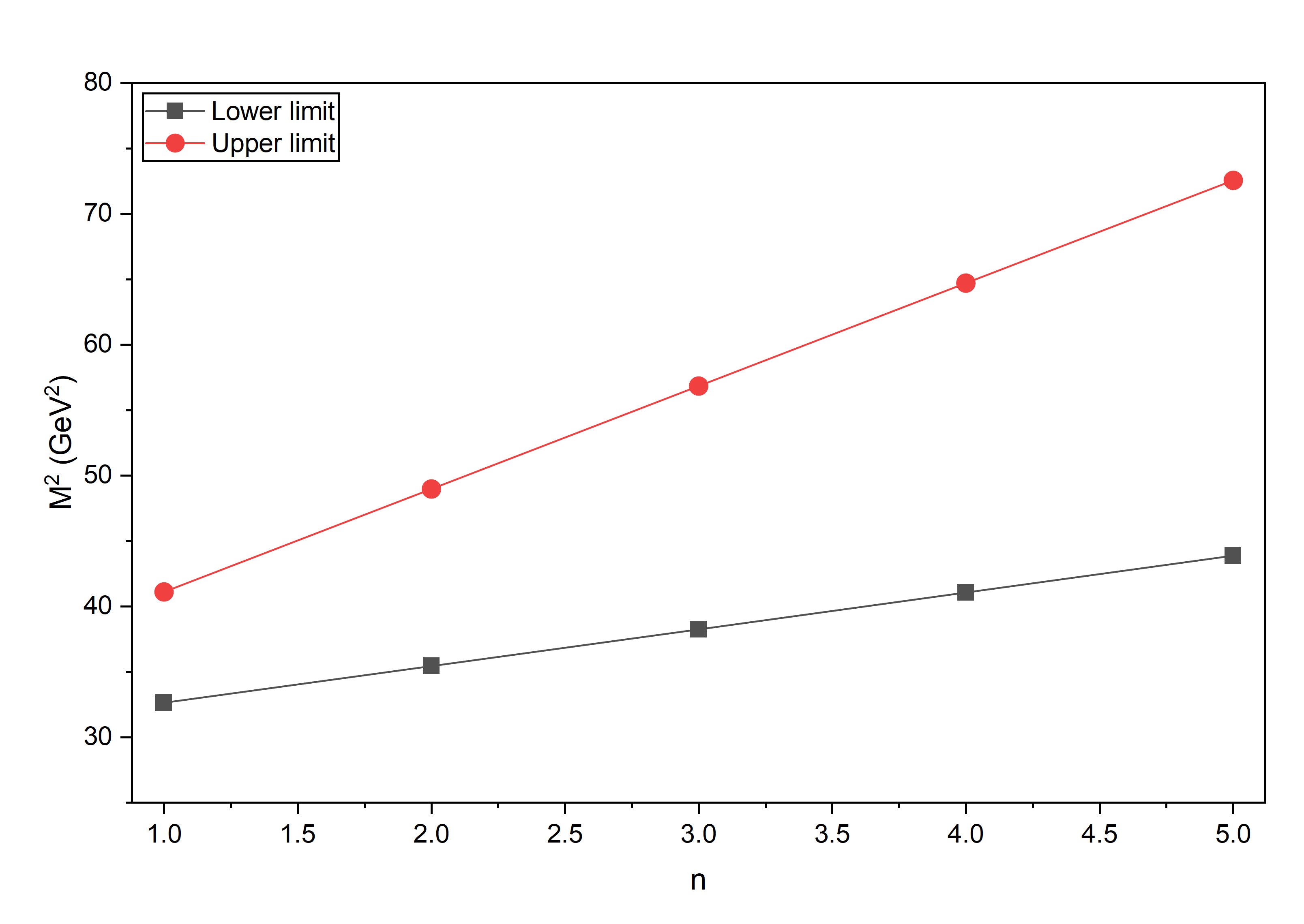}  
	\caption{Regge trajectory of $cc\bar{c}\bar{c}$ tetraqaurak for $S=0$ in $(n,M^2)$ plane.}
	\label{fig:regge trajectory for cccc of S=0 in n plane}
\end{figure}

\begin{figure}[htbp]
	\centering
	\includegraphics[width=\linewidth]{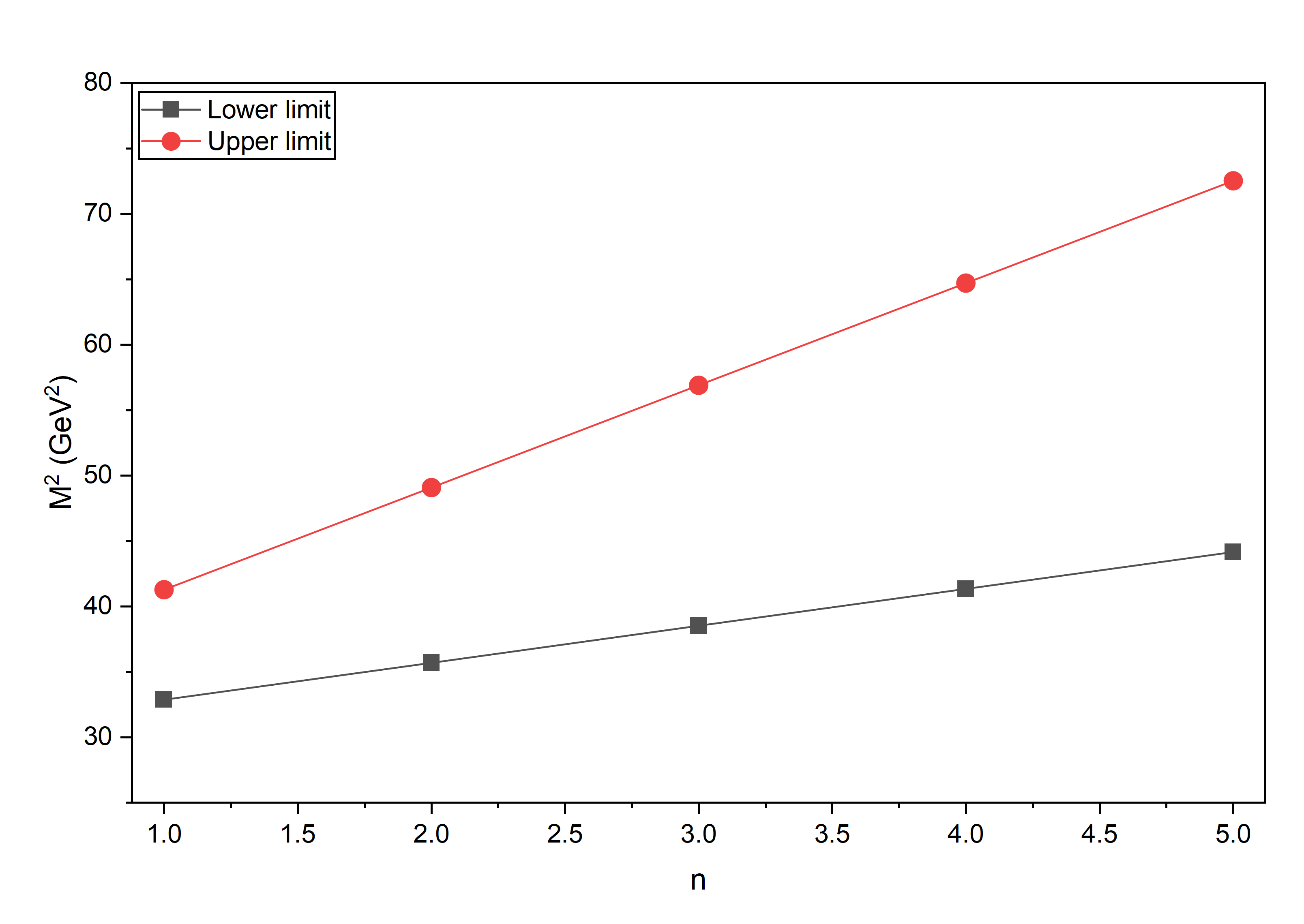}  
	\caption{Regge trajectory of $cc\bar{c}\bar{c}$ tetraqaurak for $S=1$ in $(n,M^2)$ plane.}
	\label{fig:regge trajectory for cccc of S=1 in n plane}
\end{figure}

\begin{figure}[htbp]
	\centering
	\includegraphics[width=0.95\linewidth]{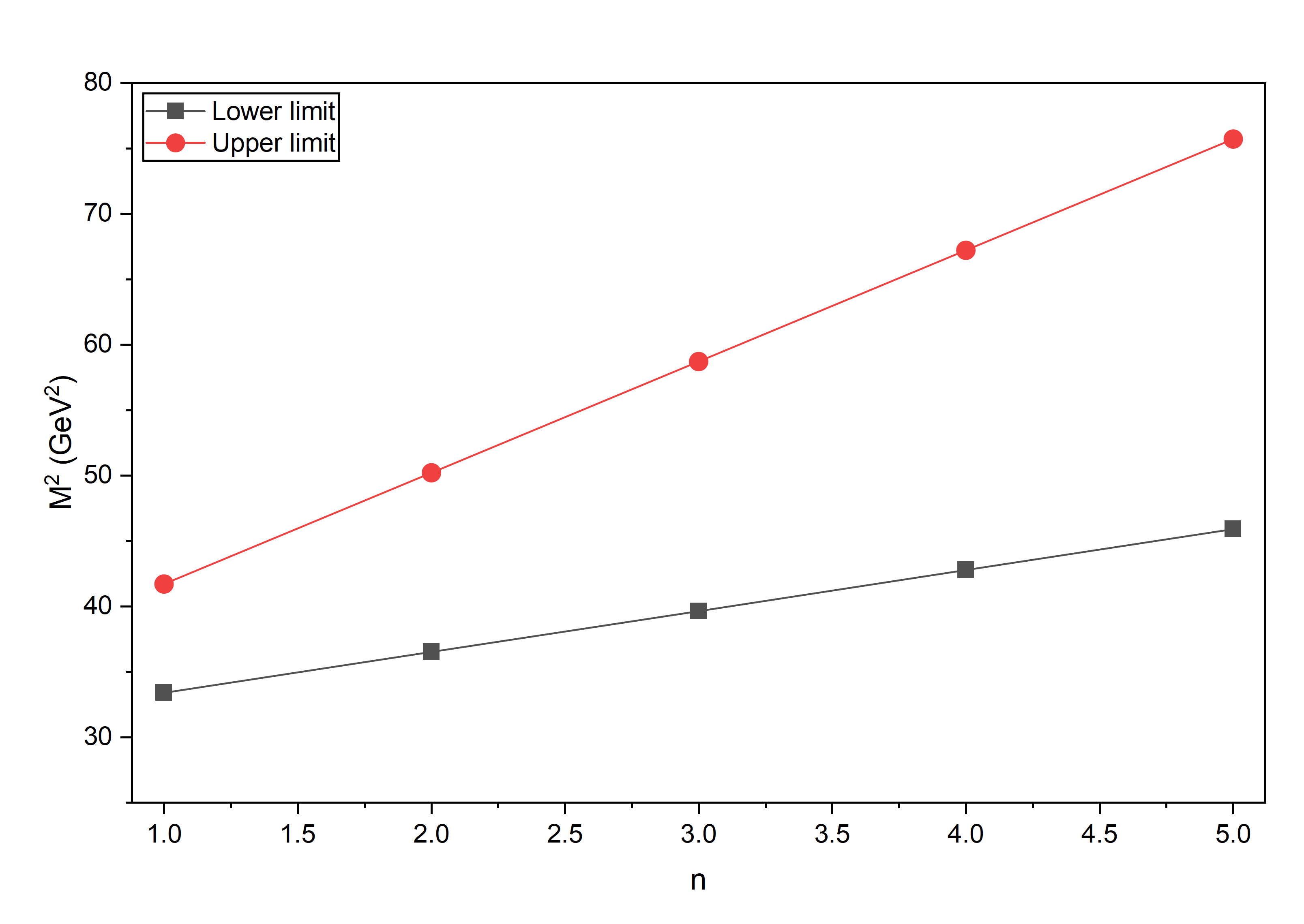}  
	\caption{Regge trajectory of $cc\bar{c}\bar{c}$ tetraqaurak for $S=2$ in $(n,M^2)$ plane.}
	\label{fig:regge trajectory for cccc of S=2 in n plane}
\end{figure}

\begin{figure}[htbp]
	\centering
	\includegraphics[width=0.95\linewidth]{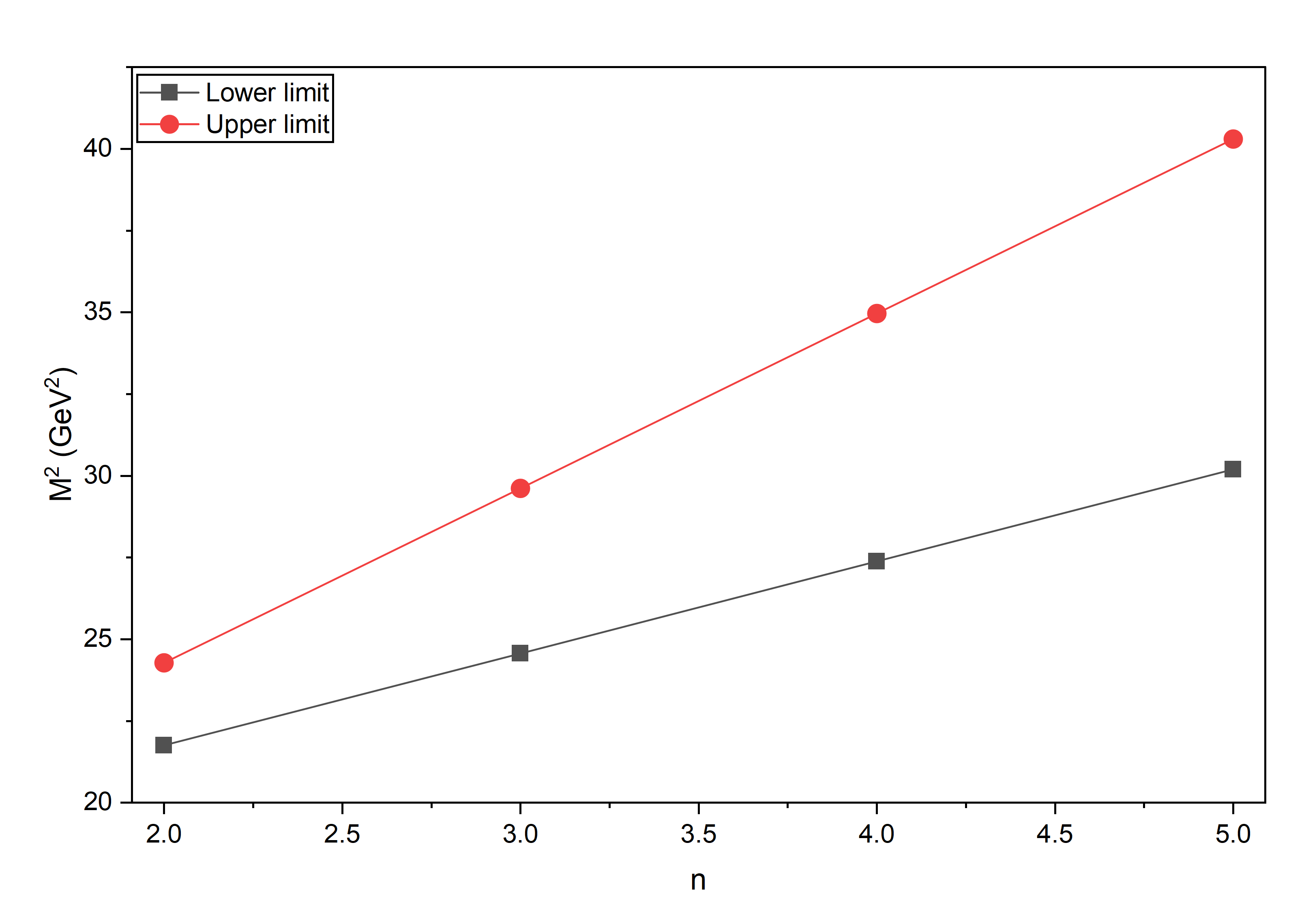}  
	\caption{Regge trajectory of $ss\bar{c}\bar{c}$ tetraqaurak for $S=0$ in $(n,M^2)$ plane.}
	\label{fig:regge trajectory for sscc of S=0 in n plane}
\end{figure}

\begin{figure}[htbp]
	\centering
	\includegraphics[width=0.95\linewidth]{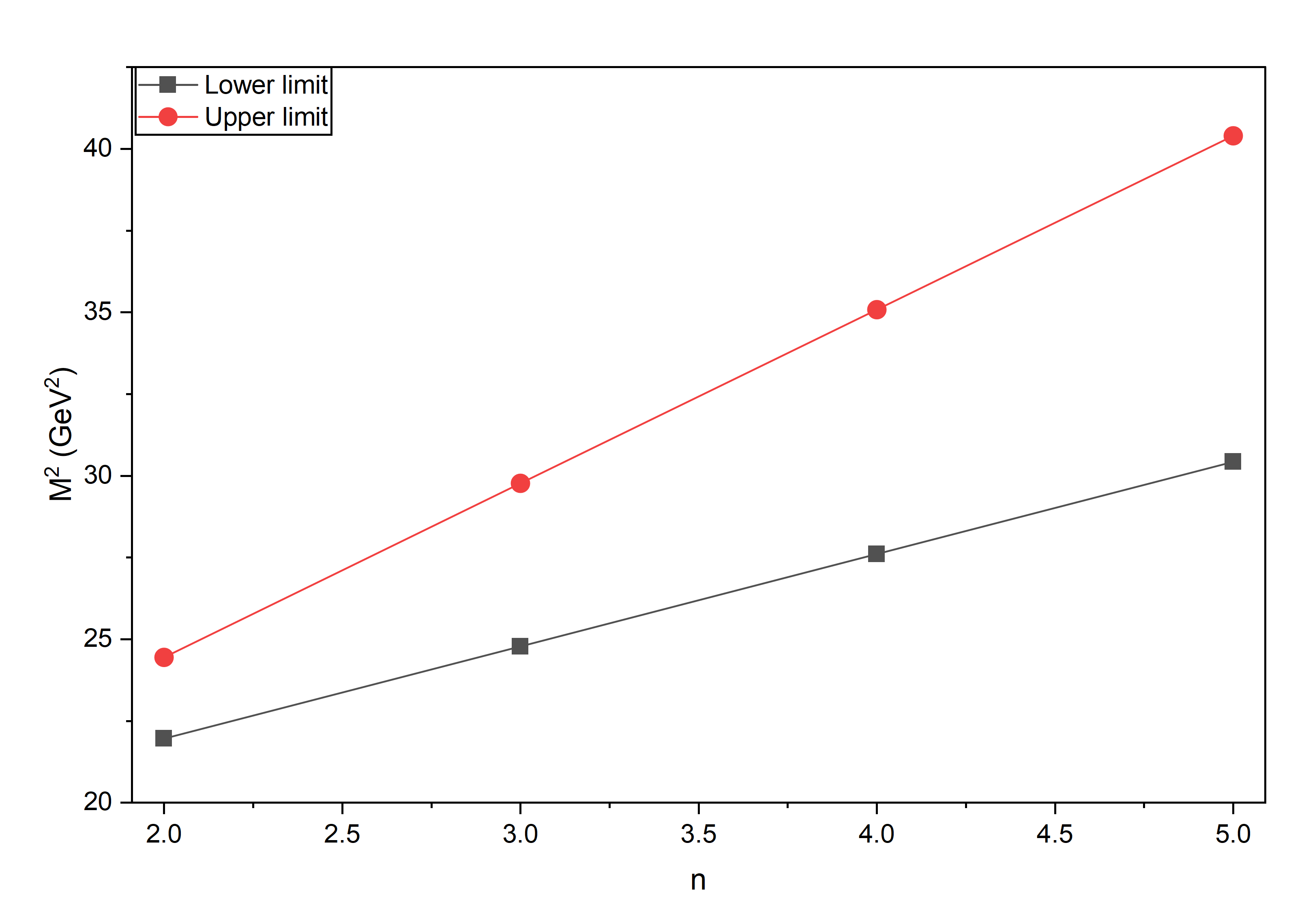}  
	\caption{Regge trajectory of $ss\bar{c}\bar{c}$ tetraqaurak for $S=1$ in $(n,M^2)$ plane.}
	\label{fig:regge trajectory for sscc of S=1 in n plane}
\end{figure}

\begin{figure}[htbp]
	\centering
	\includegraphics[width=0.95\linewidth]{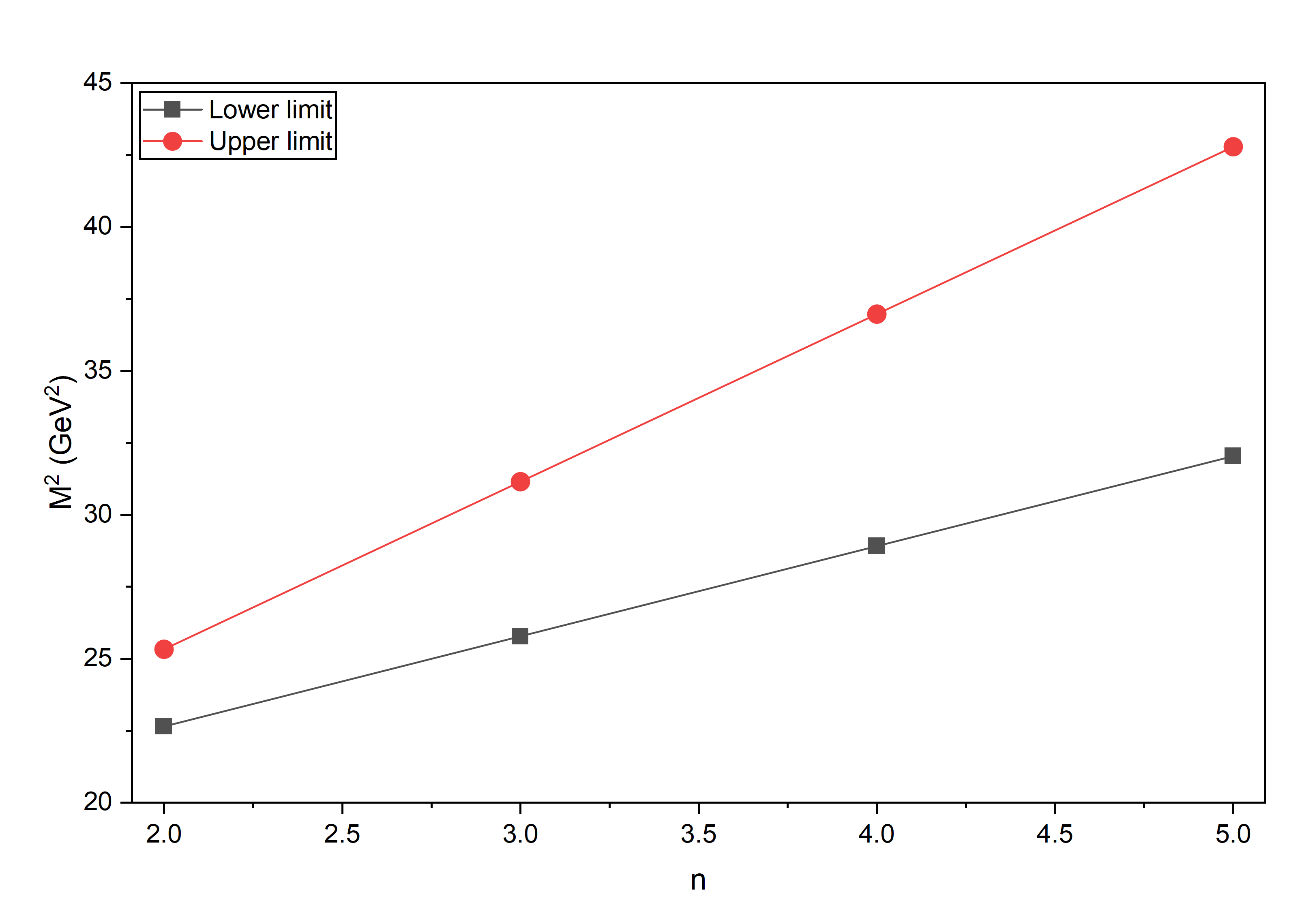}  
	\caption{Regge trajectory of $ss\bar{c}\bar{c}$ tetraqaurak for $S=2$ in $(n,M^2)$ plane.}
	\label{fig:regge trajectory for sscc of S=2 in n plane}
\end{figure}

We have also analyzed the Regge trajectories in the $(n,M^2)$ plane for the $cc\bar{c}\bar{c}$ tetraquark. These are presented in Figures~\ref{fig:regge trajectory for cccc of S=0 in n plane}, \ref{fig:regge trajectory for cccc of S=1 in n plane}, and \ref{fig:regge trajectory for cccc of S=2 in n plane}, corresponding to spin values $S=0$, $S=1$, and $S=2$, respectively. Each plot shows two trajectories: one for the lower mass limits and the other for the upper mass limits. Similarly, the Regge trajectories for the $ss\bar{c}\bar{c}$ tetraquark in the $(n,M^2)$ plane are depicted in Figures~\ref{fig:regge trajectory for sscc of S=0 in n plane}, \ref{fig:regge trajectory for sscc of S=1 in n plane}, and \ref{fig:regge trajectory for sscc of S=2 in n plane}.

\begin{table}[h]
	\caption{Values of Regge Slopes for \(ss\bar{s}\bar{s}\), \(cc\bar{c}\bar{c}\) and \(ss\bar{c}\bar{c}\) tetraquarks in \((n,M^2)\) plane (in \(\text{GeV}^{-2}\))}{\label{tab:SlopenM2}}
	\begin{tabular}{cccc}
		\toprule
		S & \(\alpha'_{ss\bar{s}\bar{s}} \, (\text{GeV}^{-2})\) & \(\alpha'_{cc\bar{c}\bar{c}} \, (\text{GeV}^{-2})\) & \(\alpha'_{ss\bar{c}\bar{c}} \, (\text{GeV}^{-2})\) \\
		\midrule
		S=0 & 0.35544 & 0.12713-0.35544 & 0.18728-0.35544 \\
		S=1 & 0.35419 & 0.12806-0.35419 & 0.18811-0.35419 \\
		S=2 & 0.31945 & 0.11763-0.31945 & 0.17194-0.31945 \\
		\botrule
	\end{tabular}
\end{table}

\begin{table}[ht]
	\centering
	\caption{Mass spectra of $cc\bar{c}\bar{c}$ tetraquark in $(n, M^2)$ plane (in GeV) and comparison with other studies}
	\begin{tabular}{ccccccc}
		\toprule
		Spin & State & $J^P$ & Calculated & Ref.~\cite{Zhao:2020} & Ref.~\cite{5Liu:2021} & Ref.~\cite{Zhao:2021} \\
		&       &       & mass (GeV) & & & \\    
		\midrule
		\multirow{5}{*}{$S=0$}
		& $1^1S_0$ & $0^+$ & $5.712$--$6.411$ & $6.346$/$6.476$ & $6.838$/$6.957$ & $6.466$ \\
		& $2^1S_0$ & $0^+$ & $5.953$--$6.998$ & $6.804$/$6.908$ & $6.954$/$7.000$/$7.183$/$6.930$ & $6.883$ \\
		& $3^1S_0$ & $0^+$ & $6.184$--$7.539$ & $7.206$/$7.296$ & $7.204$ & $7.225$ \\
		& $4^1S_0$ & $0^+$ & $6.408$--$8.044$ & -- & -- & \\
		& $5^1S_0$ & $0^+$ & $6.624$--$8.518$ & -- & -- & \\
		\midrule
		\multirow{5}{*}{$S=1$}
		& $1^3S_1$ & $1^+$ & $5.733$--$6.425$ & $6.441$ & $6.997$/$7.012$/$6.973$ & $6.494$ \\
		& $2^3S_1$ & $1^+$ & $5.974$--$7.006$ & $6.896$ & $6.934$/$7.006$ & $6.911$  \\
		& $3^3S_1$ & $1^+$ & $6.206$--$7.543$ & $7.300$ & $7.243$/$7.406$ & $7.253$ \\
		& $4^3S_1$ & $1^+$ & $6.429$--$8.044$ & -- & -- & \\
		& $5^3S_1$ & $1^+$ & $6.645$--$8.516$ & -- & -- & \\
		\midrule
		\multirow{5}{*}{$S=2$}
		& $1^5S_2$ & $2^+$ & $5.778$--$6.458$ & $6.475$ & $7.004$ & $6.551$ \\
		& $2^5S_2$ & $2^+$ & $6.043$--$7.086$ & $6.921$ & $6.942$/$7.018$ & $6.968$ \\
		& $3^5S_2$ & $2^+$ & $6.296$--$7.662$ & $7.320$ & $7.248$/$7.412$ & $7.310$ \\
		& $4^5S_2$ & $2^+$ & $6.540$--$8.198$ & -- & -- & \\
		& $5^5S_2$ & $2^+$ & $6.775$--$8.701$ & -- & -- & \\
		\midrule
	\end{tabular}
	\label{tab:cccc_spectra_innm^2}
\end{table}

\begin{table}[ht]
	\caption{Mass spectra of $ss\bar{c}\bar{c}$ tetraquark statesin $(n,M^2)$ plane (in GeV) and comparison with \cite{Tiwari:2023}.}
	\begin{tabular}{ccccc}
		\toprule
		Spin & State & $J^P$ & Calculated mass (GeV) & Ref.~\cite{Tiwari:2023} \\
		\midrule
		\multirow{4}{*}{$S=0$}
		& $2^1S_0$ & $0^+$ & 4.664--4.927 & 4.620 \\
		& $3^1S_0$ & $0^+$ & 4.956--5.442 & 4.848 \\
		& $4^1S_0$ & $0^+$ & 5.233--5.913 & -- \\
		& $5^1S_0$ & $0^+$ & 5.495--6.348 & -- \\
		\midrule
		\multirow{4}{*}{$S=1$}
		& $2^3S_1$ & $1^+$ & 4.686--4.944 & 4.638 \\
		& $3^3S_1$ & $1^+$ & 4.978--5.456 & 4.960 \\
		& $4^3S_1$ & $1^+$ & 5.254--5.923 & -- \\
		& $5^3S_1$ & $1^+$ & 5.516--6.356 & -- \\
		\midrule
		\multirow{4}{*}{$S=2$}
		& $2^5S_2$ & $2^+$ & 4.759--5.033 & 4.675 \\
		& $3^5S_2$ & $2^+$ & 5.077--5.581 & 4.972 \\
		& $4^5S_2$ & $2^+$ & 5.377--6.080 & -- \\
		& $5^5S_2$ & $2^+$ & 5.660--6.541 & -- \\
		\botrule
	\end{tabular}
	\label{tab:sscc_spectra_innm^2}
\end{table}

\section{Results and Discussion}

In this work, we apply quasi-linear Regge trajectories to map out the mass distributions of both fully-charmed ($cc\bar{c}\bar{c}$) and strange-charmed ($ss\bar{c}\bar{c}$) tetraquarks.  By fitting linear relations in the $(J,M^2)$ and $(n,M^2)$ planes, we extract trajectory slopes and intercepts for each quark configuration.  These Regge parameters then allow us to predict ground and excited state mass intervals for orbital ($J$) and radial ($n$) excitations. Below, we have discussed our calculated spectra alongside threshold values and their comparison with existing theoretical estimates.

\subsection{All-Charm ($cc\bar{c}\bar{c}$) Mass Spectra in $(J,M^2)$ plane}

Table~\ref{table:all charm_tetraquarks} lists our calculated mass ranges for all-charm tetraquarks alongside two-meson thresholds. The ground-state $1^1S_0$ ($J^P=0^+$) lies in the interval 5.712--6.411\,GeV relative to the $\eta_c(1S)\,\eta_c(1S)$ threshold at 5.968\,GeV.  Higher orbital excitations increase monotonically up to $1^5G_6$ ($6^+$) at 6.351--7.784\,GeV, each above its corresponding threshold.

To validate our results, we compare our predicted masses with those from various theoretical models and studies available in the literature~\cite{Tiwari1:2023,Wang:2017,Faustov:2021,Lundhammar:2020,Debastiani:2019,Bedolla:2020,Weng:2021,Lloyd:2004,Gordillo:2020,Berezhnoy:2012,Ader:1982,Liu:2019,Zhao:2020,Zhao:2021,RZhu:2021,Chen:2017}. The comparison is shown in Table~\ref{table:all_charm_tetraquarks_comparison}. Overall, our predicted masses show reasonable agreement with the existing theoretical predictions.  Our 1$^1S_0$ range 5.712--6.411\,GeV overlaps predictions clustering around 5.88--6.46\,GeV as shown in Table~\ref{table:all_charm_tetraquarks_comparison}.  The spin-triplet $1^3S_1$ state, predicted at 5.733--6.425\,GeV, encompasses values from 5.99 to 6.50\,GeV reported in different mentioned references.  Likewise, our spin-quintet $1^5S_2$ range 5.778--6.458\,GeV fully contains the 6.08--6.55\,GeV band given by various authors.  Predicted ranges for the $1^1P_1$ and $1^3P_2$ multiplets similarly span most existing values, and the $D$-, $F$-, and $G$-wave excitations remain within $\sim$100\,MeV of the few available estimates.  Thus, most other predicted masses lie well within our calculated windows, demonstrating the consistency of Regge phenomenology with potential models, QCD sum rules, and lattice QCD.

\subsection{$ss\bar{c}\bar{c}$ Mass Spectra in $(J,M^2)$ plane}

Table~\ref{table:sscc_tetraquarks} compiles our calculated mass ranges for $ss\bar{c}\bar{c}$ tetraquark states together with two-meson thresholds and values reported in Ref.~\cite{Tiwari:2023}.  The $1^1P_1$ ($J^P=1^-$) level is predicted at 4.542--4.706\,GeV, sitting 112--276\,MeV above the $D_s^{*\pm}D_{s0}^*(2317)^\pm$ threshold of 4.430\,GeV and in very good agreement with the 4.556\,GeV value of Ref.~\cite{Tiwari:2023}.  The $1^1D_2$ ($2^+$) state at 4.724--5.036\,GeV lies well above the $D_s^{*\pm}D_s^{*\pm}$ threshold (4.224\,GeV), while the $1^1F_3$ ($3^-$) and $1^1G_4$ ($4^+$) excitations, at 4.900--5.345\,GeV and 5.070--5.637\,GeV respectively, exceed the $D_s^{*\pm}D_{s2}^*(2573)$ (4.681\,GeV) and $D_s^{*\pm}D_{s3}^*(2860)^\pm$ (4.912\,GeV) thresholds by several MeV.

In our analysis of the mass spectra of $ss\bar{c}\bar{c}$ tetraquarks using Regge phenomenology, we find that the $1^1P_1$ state with quantum numbers $J^{PC} = 1^{--}$ lies in the mass range of 4.542--4.706\,GeV. This predicted region notably includes the experimentally observed resonance $\psi(4660)$, which has a reported mass of $4641 \pm 10$\,MeV and $J^{PC} = 1^{--}$~\cite{pdg}. The overlap between our predicted mass range and the experimental value of $\psi(4660)$, along with the matching quantum numbers, suggests a possible interpretation of this state as a $ss\bar{c}\bar{c}$ tetraquark in the $1^1P_1$ configuration. This assignment is also supported by other theoretical works, such as those based on diquark--antidiquark configurations~\cite{Tiwari:2023}, which have proposed strange-charm tetraquark interpretations for $\psi(4660)$.

In the spin-triplet sector, the $1^3P_2$ ($2^-$) interval 4.570--4.735\,GeV effectively straddles the $D_s^{*\pm}D_{s1}(2460)^\pm$ threshold (4.572\,GeV) and closely matches the 4.581\,GeV prediction of Ref.~\cite{Tiwari:2023}.  The higher $1^3D_3$ ($3^+$) and $1^3F_4$ ($4^-$) states appear at 4.757--5.071\,GeV and 4.937--5.386\,GeV, both comfortably above their respective thresholds of 4.681\,GeV and 5.320\,GeV, whereas the $1^3G_5$ ($5^+$) level at 5.111--5.683\,GeV lies in a region with no firmly established two-meson threshold.

For the spin-quintet multiplet, our $1^5P_3$ ($3^-$) mass range 4.610--4.769\,GeV exceeds the $D_s^{*\pm}D_{s2}^*(2573)$ threshold (4.681\,GeV) and agrees with 4.612\,GeV from Ref.~\cite{Tiwari:2023}.  Similarly, spin-quintet excitations span mass intervals that predominantly lie above-but in a few cases have lower bounds that approach or slightly dip below-their respective two-meson thresholds; these details are compiled in Table~\ref{table:sscc_tetraquarks}.

Overall, our predicted mass windows for $ss\bar{c}\bar{c}$ tetraquarks align closely with existing theoretical values and provide a comprehensive set of benchmarks for forthcoming experimental investigations.

\subsection{Radial Excitations of $cc\bar{c}\bar{c}$ in the $(n,\,M^2)$ Plane}

The radial spectrum of fully-charmed tetraquarks, computed via linear fits in the $(n,\,M^2)$ plane, is summarized in Table~\ref{tab:cccc_spectra_innm^2}.  For the spin-singlet sequence ($S=0$), the ground-state $1^1S_0$ interval 5.712-6.411\,GeV closely matches with Refs.~\cite{Zhao:2020,Zhao:2021}, while the first radial excitation $2^1S_0$ at 5.953-6.998\,GeV overlaps with 6.804/6.908\,GeV \cite{Zhao:2020}, 6.954-7.183\,GeV \cite{5Liu:2021} and 6.883\,GeV \cite{Zhao:2021}.  The $3^1S_0$ state in our work (6.184-7.539\,GeV) coincides with the 7.206/7.296\,GeV of Ref.~\cite{Zhao:2020} and 7.204\,GeV of Ref.~\cite{5Liu:2021}.  For higher radial levels ($n=4,5$) we extend predictions up to 8.518\,GeV, providing new targets beyond the scope of existing studies.

In the spin-triplet sector ($S=1$), our $1^3S_1$ range 5.733-6.425\,GeV agrees with 6.441\,GeV \cite{Zhao:2020} and is close to the 6.494\,GeV values for $1^3S_1$ given in Ref.~\cite{Zhao:2021}.  Subsequent excitations $2^3S_1$ and $3^3S_1$ at 5.974-7.006\,GeV and 6.206-7.543\,GeV respectively also lie within $\sim$100\,MeV of the literature.  We further predict $4^3S_1$ and $5^3S_1$ states up to $\sim$8.5\,GeV, filling out the high-$n$ spectrum.

The spin-quintet ($S=2$) radial states follow a similar pattern: the first excitation $1^5S_2$ at 5.778-6.458\,GeV is close to the 6.475\,GeV of Ref.~\cite{Zhao:2020} and 6.551\,GeV of Ref.~\cite{Zhao:2021}, while $2^5S_2$ and $3^5S_2$ (6.043-7.086\,GeV and 6.296-7.662\,GeV) encompass the 6.921-7.320\,GeV windows reported previously.  Our predictions for $n=4,5$ extend to 8.701\,GeV, charting unexplored territory for future experimental and lattice investigations.

\subsection{Radial Excitations of $ss\bar{c}\bar{c}$ in the $(n,\,M^2)$ Plane}

Table~\ref{tab:sscc_spectra_innm^2} gives the radial spectrum for strange-charmed tetraquarks in the $(n,\,M^2)$ plane.  For $S=0$, the first radial excitation $2^1S_0$ is found at 4.664-4.927\,GeV, in close agreement with 4.620\,GeV from Ref.~\cite{Tiwari:2023}, and the second excitation $3^1S_0$ at 4.956-5.442\,GeV matches the 4.848\,GeV entry.  We extend the $0^+$ predictions through $n=4,5$ up to 6.348\,GeV.

We obtain the mass range $4.664$-$4.927~\text{GeV}$ for the $2^1S_0$ state of the $ss\bar{c}\bar{c}$ tetraquark with quantum numbers $J^{PC} = 0^{++}$. This range accommodates the experimental mass of the $\chi_{c0}(4700)$ resonance, which has been reported as $4.694^{+0.016}_{-0.005}$\,GeV by the LHCb collaboration~\cite{Aaij:2016iza}. The theoretical and experimental $J^{PC}$ assignments also agree, both being $0^{++}$, which strongly supports its possible interpretation as a radial excitation of a strange-charm tetraquark. This identification is further supported by various theoretical studies. For instance, in the diquark-antidiquark model of Tiwari and Rai~\cite{Tiwari:2023}, the $\chi_{c0}(4700)$ is interpreted as a $2S$ scalar $ss\bar{c}\bar{c}$ state with $J^{PC} = 0^{++}$ and mass consistent with our predictions. Similarly, the quark delocalization color screening model (QDCSM) analysis by Liu \textit{et al.}~\cite{Liu:2021tetra} also support this assignment. The consistency between our predicted mass range and quantum numbers and those from experimental and theoretical studies lends strong support to identifying the $\chi_{c0}(4700)$ as the $2^1S_0$ $ss\bar{c}\bar{c}$ tetraquark state.

In the spin-triplet ($S=1$) channel, our $2^3S_1$ state at 4.686-4.944\,GeV aligns with 4.638\,GeV \cite{Tiwari:2023}, and $3^3S_1$ at 4.978-5.456\,GeV agrees with 4.960\,GeV.  Predictions for $n=4,5$ span 5.254-6.356\,GeV, beyond the existing data.

For the spin-quintet ($S=2$) series, the first excitation $2^5S_2$ lies at 4.759-5.033\,GeV, just above the 4.675\,GeV of Ref.~\cite{Tiwari:2023}, and the second $3^5S_2$ at 5.077-5.581\,GeV is close to the 4.972\,GeV reference.  Higher levels ($n=4,5$) are predicted between 5.377 and 6.541\,GeV, offering benchmarks for upcoming theoretical and experimental studies. Given the recent advancements in experimental facilities, such as LHCb and Belle II, the observation of such exotic states may soon become feasible, providing important tests for our theoretical predictions.

\section{Conclusion}

The Regge phenomenology applied in this work offers a remarkably economical yet powerful framework for describing both orbital and radial excitations of fully heavy and heavy-light tetraquark systems. With only few slope and intercept parameters, the model captures the overall mass ordering and level spacing without recourse to detailed dynamical potentials or lattice simulations. Its linear trajectories naturally accommodate the nearly uniformly spaced spectrum observed in our calculated points and allow straightforward extrapolation to higher spins and radial quantum numbers. The fact that our Regge-based mass windows encompass the majority of results from potential models, QCD sum rules and lattice QCD underscores the robustness of this minimal-parameter approach. Moreover, by providing continuous mass bands rather than isolated points, the Regge framework delivers clear guidance to experimental searches, suggesting where new tetraquark resonances are most likely to appear. Notably, our predicted mass ranges for the $1^1P_1$ and $2^1S_0$ $ss\bar{c}\bar{c}$ states, with $J^{PC} = 1^{--}$ and $0^{++}$ respectively, overlap well with the experimentally observed resonances $\psi(4660)$ and $\chi_{c0}(4700)$. The consistency in both mass and quantum numbers, along with support from other theoretical models, reinforces the potential interpretation of these resonances as strange-charm tetraquark states. In this way, our study demonstrates that Regge phenomenology remains a vital tool for organizing and predicting the spectroscopy of exotic multiquark states.

Thus, our results contribute to the growing theoretical efforts aimed at understanding the spectroscopy of exotic tetraquark states, particularly the fully-heavy systems like $cc\bar{c}\bar{c}$ and heavy-light systems like $ss\bar{c}\bar{c}$, and serve as valuable references for future experimental and theoretical investigations.

\FloatBarrier

\section{Acknowledgment}
Vandan Patel acknowledges the financial assistance by
University Grant Commision (UGC) under the CSIR-UGC
Junior Research Fellow (JRF) scheme with Ref No.
231610186052.

\end{document}